\documentclass[journal]{IEEEtran}

\ifCLASSINFOpdf
\else
\fi
\usepackage{amssymb}
\usepackage{mathptmx} 
\usepackage{array}
\usepackage{fancyhdr}
\usepackage[normalem]{ulem}
\usepackage[hyphens]{url}
\usepackage[sort,nocompress]{cite}
\usepackage[final]{microtype}
\usepackage[keeplastbox]{flushend}
\usepackage[bookmarks=true,breaklinks=true,letterpaper=true,colorlinks,linkcolor=black,citecolor=blue,urlcolor=black]{hyperref}
\usepackage{setspace}
\usepackage[framemethod=tikz]{mdframed}
\usepackage{framed}
\usepackage{color}
\definecolor{shadecolor}{rgb}{0.94,0.96,1.0}

\usepackage{amsmath}
\usepackage{setspace} 
\newcommand{\myfont}{\fontsize{9pt}{\baselineskip}\selectfont}
\usepackage[linesnumbered, ruled, boxed]{algorithm2e}
\usepackage{framed}
\usepackage{listings}
\usepackage{subfigure}
\usepackage{multirow}
\usepackage{xcolor}  
\usepackage{amsmath}
\usepackage{graphics}
\usepackage{booktabs}
\usepackage{pifont}
\usepackage{ulem}
\usepackage{tikz}
\newcommand*{\circled}[1]{\lower.7ex\hbox{\tikz\draw (0pt, 0pt)%
    circle (.5em) node {\makebox[1em][c]{\small #1}};}}
\usepackage{titlesec}
\makeatletter
\g@addto@macro{\normalsize}{%
  \setlength{\abovedisplayskip}{3pt plus 0.5pt minus 1pt}
  \setlength{\belowdisplayskip}{3pt plus 0.5pt minus 1pt}
  \setlength{\abovedisplayshortskip}{0pt}
  \setlength{\belowdisplayshortskip}{0pt}
  \setlength{\intextsep}{4pt plus 1pt minus 1pt}
  \setlength{\textfloatsep}{4pt plus 1pt minus 1pt}
  \setlength{\skip\footins}{5pt plus 1pt minus 1pt}
  }
\makeatother

\usepackage{titlesec}
\usepackage{pifont}
\usepackage{multirow}

\titlespacing\section{0pt}{4pt plus 1pt minus 1pt}{5pt plus 1pt minus 2pt}
\titlespacing\subsection{0pt}{4pt plus 1pt minus 1pt}{5pt plus 1pt minus 2pt}
\titlespacing\subsubsection{0pt}{2pt plus 1pt minus 1pt}{3pt plus 1pt minus 2pt}

\begin{document}
%
\title{Energon: Towards Efficient Acceleration of Transformers Using  Dynamic Sparse Attention}

\author{Zhe~Zhou,
        Junlin~Liu,
        Zhenyu~Gu, Guangyu~Sun,~\IEEEmembership{Member,~IEEE}
\thanks{Z.Zhou, J.Liu and G.Sun are with the Center for Energy-efficient Computing and Applications, Peking University, 100871, Beijing, China. (e-mail:
gsun@pku.edu.cn)}
\thanks{Z.Gu was with Alibaba Inc. Beijing, China. (e-mail: guzhenyu@gmail.com)}
\thanks{(Corresponding author: Guangyu Sun)}
}

%



\newcommand\trev[1]{#1}
\newcommand\rev[1]{{#1}}
\newcommand\todo[1]{\textcolor{red}{#1}}

\newcommand\ServerGeomeanSpeedupOverXeon{\trev{$168\times$}}
\newcommand\ServerGeomeanSpeedupOverGPU{\trev{$8.7\times$}}
\newcommand\SpeedupOverSpAtten{{$1.7\times$}}
\newcommand\SpeedupOverAcube{{$1.25\times$}}
\newcommand\EnergyOverSpAtten{{$1.6\times$}}
\newcommand\EnergyOverAcube{{$1.5\times$}}

\newcommand\EdgeGeomeanSpeedupOverTX{\trev{$37\times$}}
\newcommand\EdgeGeomeanSpeedupOverARM{\trev{$440\times$}}
\newcommand\EdgeGeomeanEnergyOverTX{\trev{$340\times$}}
\newcommand\EdgeGeomeanEnergyOverARM{\trev{$2951\times$}}

\maketitle

\begin{abstract}

In recent years, transformer models have revolutionized Natural Language Processing (NLP)  and  shown promising performance on Computer Vision (CV) tasks.
Despite their effectiveness, transformers' attention operations are hard to accelerate due to the complicated data movement and quadratic computational complexity, prohibiting the real-time inference on resource-constrained edge-computing platforms.

To tackle this challenge, we propose Energon, an algorithm-architecture co-design approach that accelerates various transformers using dynamic sparse attention. 
With the observation that  attention results only depend on a few important query-key pairs, we propose a \trev{\emph{Mix-Precision Multi-Round Filtering (MP-MRF)}} algorithm to dynamically identify such pairs at runtime. We adopt low bitwidth in each filtering round and only use high-precision tensors in the  attention stage to reduce overall complexity. By this means, we significantly  mitigate the computational cost with  negligible accuracy loss. To enable such an algorithm with lower latency and better energy efficiency, 
we also propose an Energon co-processor architecture. Elaborated pipelines and specialized  optimizations jointly boost the performance and reduce power consumption. Extensive experiments on both NLP and CV benchmarks demonstrate that {Energon  achieves \ServerGeomeanSpeedupOverXeon~and \ServerGeomeanSpeedupOverGPU~geo-mean speedup and up to $10^4\times$ and $10^3\times$ energy reduction over  Intel    Xeon 5220 CPU  and NVIDIA V100 GPU, respectively }    Compared to state-of-the-art attention accelerators SpAtten and $A^3$, Energon also achieves  \SpeedupOverSpAtten,\SpeedupOverAcube~speedup  and  \EnergyOverSpAtten,\EnergyOverAcube~higher energy efficiency.
\end{abstract}


%
\IEEEpeerreviewmaketitle




\section{Introduction}

\IEEEPARstart{B}{enefiting} from the powerful attention mechanism,   transformer  models 
such as Seq2seq~\cite{transformer}, BERT~\cite{bert}, GPT-2~\cite{gpt2},  XLNet~\cite{xlnet}, T5~\cite{t5} and other variants~\cite{albert,roberta,bart}
have achieved leading-edge performance on various NLP tasks, such as  question-answering, text classification and  machine translation, etc.  Besides, transformers also show promising performance on many computer vision tasks including image classification~\cite{vit, deit}, object detection~\cite{detr, deformable} and even video comprehension~\cite{videobert, video_transformer}. In brief, transformers are becoming the de facto substitutes for traditional RNNs and CNNs in a broad range of  scenarios.

\begin{figure} [t]
\label{bcm_computation}
\setlength{\abovecaptionskip}{-2pt} 
    \centering
    \includegraphics[width=0.98\linewidth]{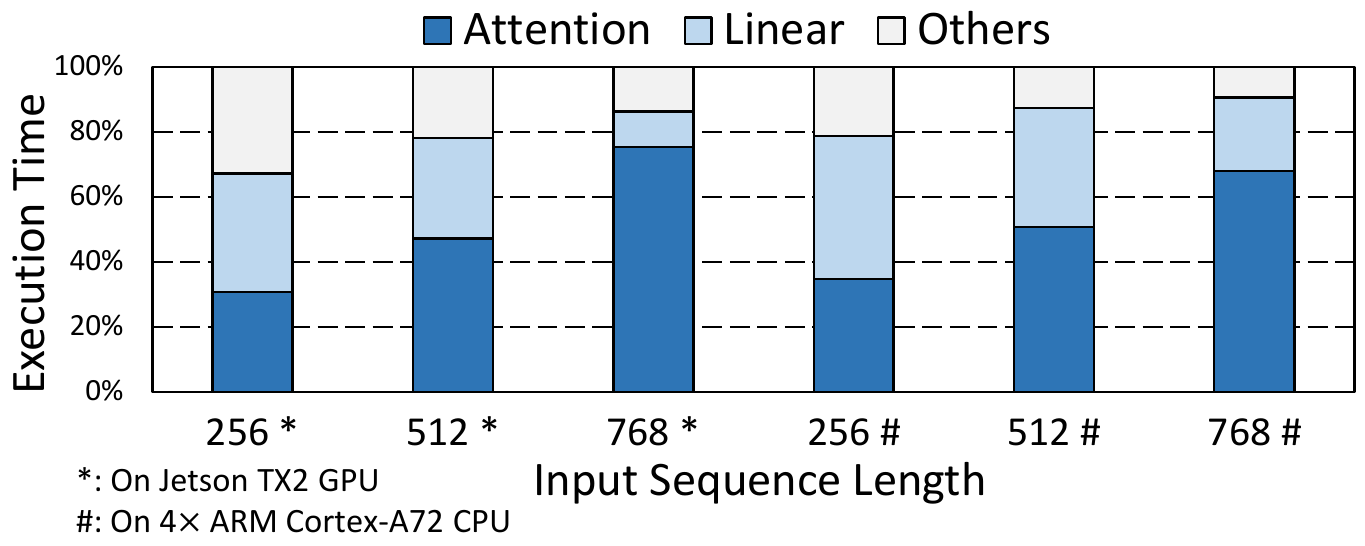} 
      \caption{Execution time breakdown of the BERT-base model on two typical edge-computing platforms. 
      }
            \label{fig:profiling}
\end{figure}

Despite their effectiveness, it is still challenging to deploy transformers on resource-constrained  devices. The greatest difficulty comes from 
the attention operations, which involve complicated data movement \cite{a3, spatten} and  bear quadratic computational  complexity concerning the input sequence length~\cite{reformer, longformer, routing_transformer,blockbert}. Therefore, for many tasks having long input sequences like question-answering and image-classification, it is always  difficult to realize low-latency inference.  
For instance, we have  profiled the execution time of the BERT-base model~\cite{bert} on two popular edge-computing  platforms. As shown in   Figure~\ref{fig:profiling}, the  attention operations take up about half of
the total execution time as the sequence length grows to $512$. When the input length reaches $768$,  attention operations even dominate the total execution (70\% on average).
Note that the attention overhead is becoming more prominent as weight-pruning and quantization  methods have significantly  reduced the complexity of linear  layers in transformers~\cite{ternaryBert, gobo,q8bert,qbert}. 
Specialized NN accelerators have also been  proposed to handle the non-attention parts~\cite{eie, diannao, gobo}. Therefore, how to process attention operations efficiently is the key problem of  transformers acceleration.

Some recent works propose to co-design  attention algorithms and accelerator architectures to mitigate the attention overhead. They mainly exploit the sparsity of attention to speed up  the execution. For instance, $A^3$~\cite{a3} leverages several approximation  strategies to avoid computing near-zero scores to reduce the computational overhead.  SpAtten~\cite{spatten}  proposes a cascaded token pruning mechanism that progressively prunes unimportant tokens to reduce the overall complexity.   However, these two solutions still have intrinsic drawbacks. Specifically, $A^3$ needs to load all the data on-chip to perform approximate computation, failing to reduce off-chip DRAM access. The cascaded token pruning used in SpAtten successfully reduces both computation and DRAM access, but it is a coarse-grained strategy that  does not support  dynamic pruning for different attention heads.  Our experiments reveal that token pruning can hardly achieve acceptable accuracy without model retraining.   Moreover, previous works only conduct evaluations  on NLP tasks. Their  performance on  vision tasks, e.g., image classification, remains to be explored.

\renewcommand{\thefootnote}{\fnsymbol{footnote}}
\begin{figure} [t]
    \centering
    \includegraphics[width=0.9\linewidth]{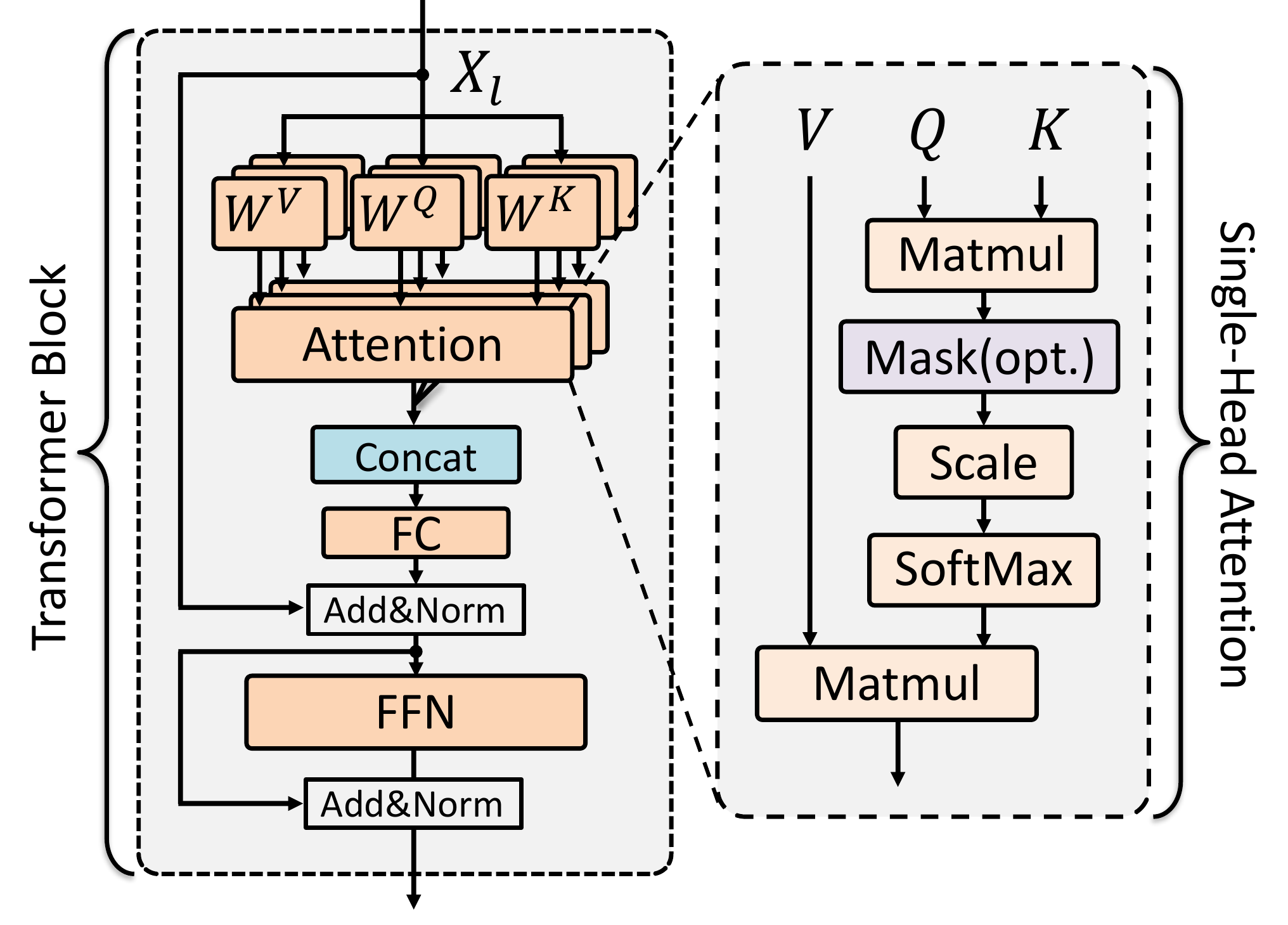} 
    \vspace{-0.5em}
      \caption{The standard transformer block. \trev{$W^Q,W^K,W^V$ are projection weights to project the input feature $X_l$ to Query ($Q$), Key ($K$), and Value ($V$) features}. }
            \label{fig:transformer}
\end{figure}

To overcome these problems,  we propose Energon\footnote{In the famous film series \emph{Transformers}, Energon  is the preferred fuel of the Transformer race.}, an algorithm-architecture co-design solution that efficiently accelerates transformers using dynamic sparse attention. With the observation that attention results mainly depend on a few important query-key pairs, we propose a novel \trev{\emph{Mix-Precision Multi-Round Filtering (MP-MRF)} } algorithm to  identify these  pairs at runtime. We adopt  low-bitwidth operations  in each filtering round.  Only the finally selected pairs are used for sparse attention with high-precision tensors. By this means, we reduce $4\times$ to $8\times$ computation with negligible accuracy loss.  To enable such an algorithm with lower latency and better energy efficiency, 
we also design a co-processor architecture named Energon. It boosts the performance and saves DRAM access through elaborated pipelines and several  optimization strategies. In addition, we propose a performance model to analyze the bottlenecks of Energon architecture under different situations so as to guide the hardware configuration. 
Extensive experiments on both NLP and CV benchmarks show that, on average, Energon gets \rev{\EdgeGeomeanSpeedupOverTX~and \EdgeGeomeanSpeedupOverARM} geo-mean speedup and achieves \rev{\EdgeGeomeanEnergyOverTX~and \EdgeGeomeanEnergyOverARM} geo-mean  energy reduction  compared to TX2 GPU and ARM-A72 CPU, respectively. 
Besides, to meet the requirements in cloud-computing scenarios, we also present an Energon-server which shows \rev{ \ServerGeomeanSpeedupOverXeon, \ServerGeomeanSpeedupOverGPU} geo-mean speedup, and  \rev{ $10^4 \times $,$10^3 \times $ }energy-efficiency compared to Intel Xeon CPU and  NVIDIA-V100 GPU. We further compare our design against the  state-of-the-art attention accelerators and prove that Energon achieves on-average \SpeedupOverSpAtten,\SpeedupOverAcube~speedup and \EnergyOverSpAtten,\EnergyOverAcube~energy-efficiency over SpAtten and  $A^3$. To summarize, we have the following main contributions:

\begin{itemize}
    
    \item We propose a mix-precision multi-round filtering (MP-MRF) algorithm to realize dynamic sparse attention in transformers (Section~\ref{sec:algorithm}). Without retraining, it achieves \trev{$4\times$ to $8\times$} pruning ratio  with negligible accuracy loss  on both NLP and CV tasks (Section~\ref{sec:eval1}).  
    \item We design an Energon architecture to accelerate the proposed algorithm.  It  functions as a co-processor that is plug-in compatible with other NN accelerators. We carefully design the pipelines and propose several architectural optimizations to  boost the performance and reduce energy consumption. 
    A performance model is also introduced to help determine the hardware configurations (Section~\ref{sec:accelerator}).
    \item We 
    prove that Energon  achieves multiple orders of magnitude speedup and energy reduction over commodity CPUs/GPUs. Energon also outperforms state-of-the-art attention accelerators SpAtten and $A^3$ (Section~\ref{sec:eval}).

\end{itemize}


\section{Background and Motivation}

\label{sec:background}

\subsection{Transformer Algorithms }
\label{sec:transformer_algorithm}

\begin{algorithm}[t]
\label{algo:1}
\caption{Multi-Head-Attention}
\textbf{Input:} $\{Q,K,V\}\in \mathbb{R}^{n\times d}$, number of heads $H$; \\
Split $Q,K,V$ to $H$ chunks respectively; \\
$d_h = d/H$;\\

\For{head\_index h~$\leftarrow$~\emph{0} until $H$}{
\emph{attention\_score} $\leftarrow$ $Q_h\cdot K_h^T/\sqrt{d_h}$;\\
\emph{attention\_score} $\in \mathbb{R}^{n\times n}$; \\ 
\For{row\_index i~$\leftarrow$~\emph{0} until $n$}{
    \emph{attention\_prob}[$i$] $\leftarrow$ Softmax(\emph{attention\_score}[$i$]); \\ 
    }
    \emph{attention\_result}[$h$] $\leftarrow$ \emph{attention\_prob}$\cdot V_h$;
}
\emph{output} $\leftarrow$ concat(\emph{attention\_result});\\
\textbf{Return:} \emph{output $\in  \mathbb{R}^{n\times d}$};
\end{algorithm}

Transformers, first proposed by Google~\cite{transformer}, have demonstrated leading-edge performance on a variety of NLP tasks, including discriminative language understanding tasks~\cite{bert,roberta,albert} and generative language modeling tasks~\cite{transformer,t5,gpt2}. Surprisingly,  transformers are also gaining more and more popularity among computer vision tasks~\cite{deformable, detr,vit,botnet,video_transformer}. Thanks to their outstanding effectiveness and generality, transformers are considered substitutes for traditional RNNs and CNNs in many real-world scenarios.

Unlike traditional DNNs~\cite{alexnet, vgg,  rnn}, transformer models are generally built by stacking several  transformer blocks. As illustrated in Figure~\ref{fig:transformer}, a standard transformer block consists of  three key components, namely Linear layers \trev{(including projection layers, fully-connected layers (FC) and the feed-forward network (FFN))}, Multi-Head Attention ~(Attention), and Normalization layers~(Norm).   
For the $l$-th block of a transformer model, the input is a sequence of $n$ vectors (tokens), denoted as $X_l\in \mathbb{R}^{n\times d_{in}}$ where $n$ and $d_{in}$ are the sequence length and the \trev{input feature dimension}. Three linear projection weights $\{W^Q, W^K, W^V\} \in \mathbb{R}^{d_{in}\times d}$ project the input  $X_l$ to Query, Key and Value, denoted as \{$Q$, $K,$ $V\} \in \mathbb{R}^{n\times d}$, \trev{where $d$ is the hidden feature size}.  Attention is then performed on these features to capture long-term dependencies of the input sequence. 

As described in Algorithm~\ref{algo:1}, an attention layer has in total $H$ different attention heads, each processing a \trev{chunk} of $Q,K,V$. Therefore, we first split $Q,K,V$ to $H$ chunks, respectively. Single-head attention is then applied to each chunk. Just as illustrated in the right part of Figure~\ref{fig:transformer}, the $h$-th head  computes the outputs as follows:
\begin{equation}
    \emph{Attention}(Q_h,K_h,V_h) = \mathit{Softmax}(\frac{Q_h\cdot K_h^\intercal}{\sqrt{d_h}})V_h
\end{equation}

\noindent The \emph{attention score}, namely the  scaled dot-production result of query $Q_h$ and key $K_h$, is passed to a \emph{Softmax} function to get the \emph{attention probability} \trev{(we use $d_h$ to denote the feature size of one head chunk)}. Then the probability matrix  is multiplied with value tensor $V_h$ to derive the single-head attention result, \trev{which has a shape of $n\times d_h$}.
Each head computes \emph{attention result} individually. The results are concatenated to form the final results.  \trev{Note that  in the rest of this paper, } we also use $Q,K,V$ to denote the single-head inputs $Q_h, K_h, V_h$ and use $d$ to represent a single head's feature dimension for simplicity.
Apart from  Multi-Head-Attention, the remaining operations like   Normalization and FFN  have been widely applied in traditional DNNs. To be specific, the FFN  has two fully-connected layers:
\begin{equation}
    \text{FFN}(X_l) = \text{GELU}(X_l W_1+b_1) W_2 +b_2
\end{equation}
\vspace{-1pt}

\noindent where $W$ and $b$ represent the weight and bias, and GELU is a special activation function~\cite{gelu}.

 Since the attention layer conducts pair-wise dot-productions among $n$ vectors, the complexity should be $O(n^2d)$, while linear layers' complexity is about $O(nd^2)$, because they process each input token individually. Given that $d$ is usually a  fixed value, the overhead of attention operations grows more rapidly than the linear parts as the length of input sequence $n$ increases. Our profiling results in Figure~\ref{fig:profiling} have vividly shown this trend. On both embedded GPU and CPU platforms, the attention operations take up more than half of the total execution time as  $n$ increases larger than 512 and dominate the execution time when $n$ is set to 768. That is to say, for tasks requiring long input sequences, the attention operations will become the bottleneck. Moreover, the \trev{challenge of attention acceleration} is becoming increasingly prominent as weight pruning~\cite{deepcompression}, quantization~\cite{gobo,q8bert} and specialized NN accelerators~\cite{eyeriss, eie, diannao, gobo} can provide efficient acceleration of linear layers in transformers.
Therefore, it is urgent to devise solutions to execute attention operations more efficiently.



\subsection{Sparse Attention}

\label{sec:sparse}

Several algorithms and accelerators focused on exploring the sparsity in attention layers have been proposed  to mitigate the attention overhead of transformers.  As described in Algorithm \ref{algo:1}, the   dot-production of $Q$ and $K$ generates the $n\times n$ attention score matrix.  Since a \emph{Softmax} function is applied to each row of the score matrix to derive the attention probability, the chances are that only a small fraction of scores in each row will produce a large probability, and most of the dot-production scores are too small to affect the final attention results. 

\begin{figure} [t]
    \centering
    \includegraphics[width=1.0\linewidth]{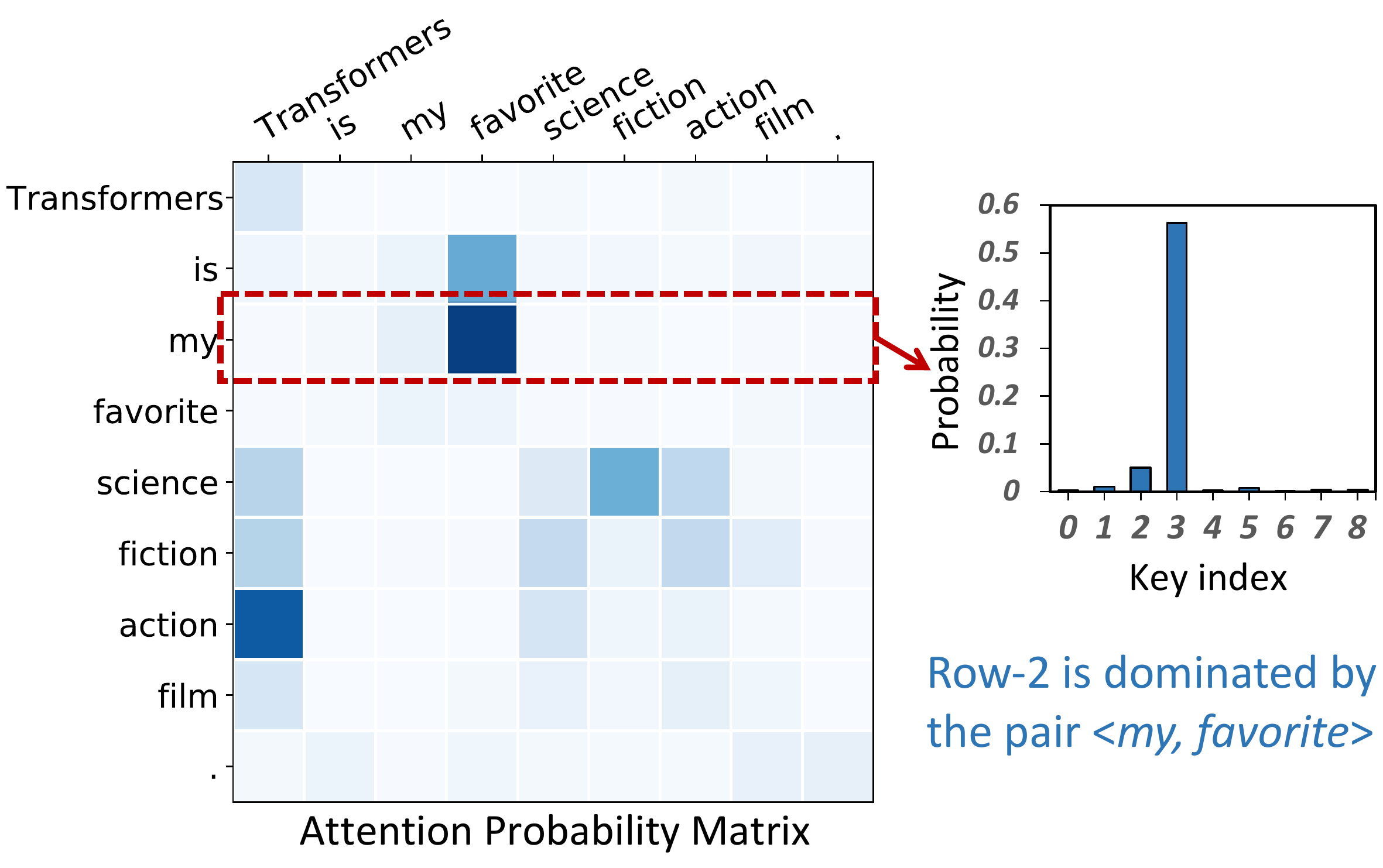} 
      \caption{Sparsity in attention probability matrix. \trev{In the annotated row, the data distribution is dominated by the query-key pair $<$\emph{my, favorite}$>$. }}
            \label{fig:probability}
\end{figure}

Figure~\ref{fig:probability} illustrates one of the  probability matrices extracted from the BERT-base model trained on the SST-2 dataset\cite{sst2}. For most of the rows, 
only a small portion of elements have a large probability. For instance, in the annotated row, the probability of pair $<$\emph{my, favorite}$>$ dominates the distribution.
If we only use those ``important" query-key pairs for attention and ignore the others, we can save much computation. Moreover, we can only load the selected keys and  values on-chip to reduce total memory access. Therefore, the question becomes: \emph{How to find the important query-key pairs?} Previous works'  answers to this problem have formed two trends: content-independent sparsity and content-based sparsity.

\textbf{Content-independent sparsity}: A lot of prior works simply force the attention matrix to be sparse through applying fixed, predefined patterns such as local windows~\cite{sparse}, strided patterns \cite{sparse,longformer}, blockwise patterns~\cite{blockbert} or combining two or more distinct patterns~\cite{ho2019axial,sparse}.  However, these heuristic methods fail to  find the important query-key pairs  explicitly since they do not take the input features' content into consideration.  Therefore, the performance is always limited~\cite{efficient}.

\textbf{Content-based sparsity}: Unlike content-independent methods, content-based methods dynamically determine the sparsity pattern through analyzing the inputs. To name a few,  Reformer~\cite{reformer}  uses local-sensitive-hashing (LSH) to select the query-key pairs that are close in the hashing space. It reduces the attention complexity to $O(n\log n)$.
Meanwhile, Routing-transformer~\cite{routing_transformer}  leverages clustering algorithms to find the important pairs before attention. It reduces the computational complexity from $O(n^2d)$ to $O(n^{1.5}d)$.    However, both methods have intrinsic drawbacks: The LSH hashing needs to be performed multiple rounds (up to 8 according to the experiments) to guarantee accuracy, 
and
the clustering algorithm itself is too complex for embedded accelerators. Some other works~\cite{Sinkhorn,informer} also propose dynamic sparse attention methods. However, they only focus on algorithm improvements and seldom consider hardware designs.

\subsection{SW/HW Co-Design Approaches}

We notice that a few previous works attempt to accelerate attention operations through software-hardware co-design. As mentioned before, $A^3$~\cite{a3} accelerates attention operations in neural networks with algorithmic approximation and hardware specialization.  However, $A^3$ only saves computation and cannot reduce DRAM access. Another recent work  
SpAtten~\cite{spatten} proposes a cascaded token pruning mechanism to prune unimportant tokens layer by layer, which reduces both computation and DRAM access. However,  token-pruning does not support fine-grained dynamic pruning for different attention heads. According to our evaluation, it fails to achieve acceptable accuracy in the absence of retraining.  Moreover, these two works are only evaluated on NLP tasks. Whether sparse attention works on CV tasks remains to be discussed.

\section{Energon Algorithm}
\label{sec:algorithm}


\label{sec:filtering}

We argue that an ideal sparse attention mechanism should meet three following goals:
(1) It should achieve a high pruning ratio and maintain the models' accuracy without retraining, as the training dataset is  always inaccessible in deployment scenarios. (2) It should be compatible with various transformer models, including NLP  and CV models. (3) It should be hardware-friendly and have the potential to reduce both computation and DRAM access.
With these goals, we propose a  dynamic sparse attention algorithm using a novel multi-round filtering strategy. Details are introduced as follows.

\subsection{Top-k Pruning: A Baseline}

\begin{figure} [t]
    \centering
    \includegraphics[width=1.0\linewidth]{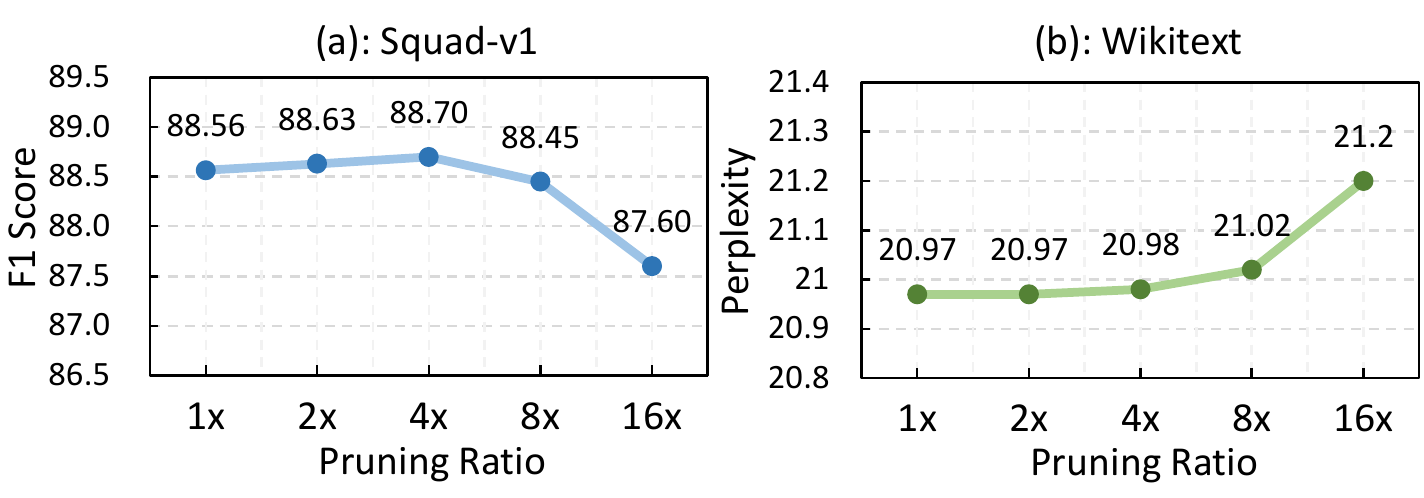} 
    \vspace{-1.5em}
      \caption{Accuracy  VS. pruning ratio using Top-k pruning. \trev{(a): Question-answering task on Squad-v1 dataset.  (b): Language-modeling task on Wikitext. }}
            \label{fig:pruning_ratio}
\end{figure}

We begin our discussion with a question: \emph{How to identify the query-key pairs that will produce large attention probabilities at runtime?} 
Given an attention score matrix, a straightforward way is to keep the $k$-largest elements in each row to calculate the probability.  We view this method as the baseline solution. To explore the efficiency of top-k pruning,  we evaluate it on two representative tasks: (a) question-answering task on the SQuAD-v1 dataset~\cite{squad} using the BERT-base model~\cite{bert}. (b) language-modeling task on the Wikitext-2 dataset~\cite{wikitext} using the GPT-2 model~\cite{gpt2}. The evaluation code and pre-trained models are provided by the popular Hugging-Face \trev{Transformers}~\cite{hugging_face} project. For each row of the attention score matrices, we sort and select the $k$-largest scores to conduct \emph{Softmax}. The  unselected elements are set to zero.  We follow SpAtten's~\cite{spatten} practice and do not prune the first two blocks, \trev{since the first few blocks have not captured high-level relations between tokens and therefore do not show obvious sparse patterns}. Without any retraining, the accuracy (F1-score for task (a), higher is better; Perplexity for task (b), lower is better) under different pruning ratios are plotted in Figure~\ref{fig:pruning_ratio}. As we can see, when 87.5\% of the total keys of each row are pruned ($8\times $ pruning ratio), the two tasks only bear negligible accuracy loss (-0.12\% for SQuAD and +0.05 for Wikitext). Even with a 16$\times$ pruning ratio, the accuracy loss is within 1\%. Such results demonstrate the huge potential of leveraging the sparsity of attention matrices to reduce the attention overhead.

However, the na\"ive top-k pruning has two obvious drawbacks: First, the pruning is conducted after the calculation of attention scores, which only saves \emph{Softmax} and $\emph{attention\_prob}\times V$ computation. We still have to conduct the $O(n^2)$ scaled dot-production, namely $Q\cdot K^T/\sqrt{d}$. Second, top-k pruning requires specialized sorting engines to select the $k$-largest elements. For example, both $A^3$ and  SpAtten spend much effort designing the top-k engines. We argue that a simpler method is preferred to help select the important query-key pairs.





\subsection{\trev{Mix-Precision} Multi-Round Filtering }

\begin{figure} [t]
    \centering
    \includegraphics[width=1.0\linewidth]{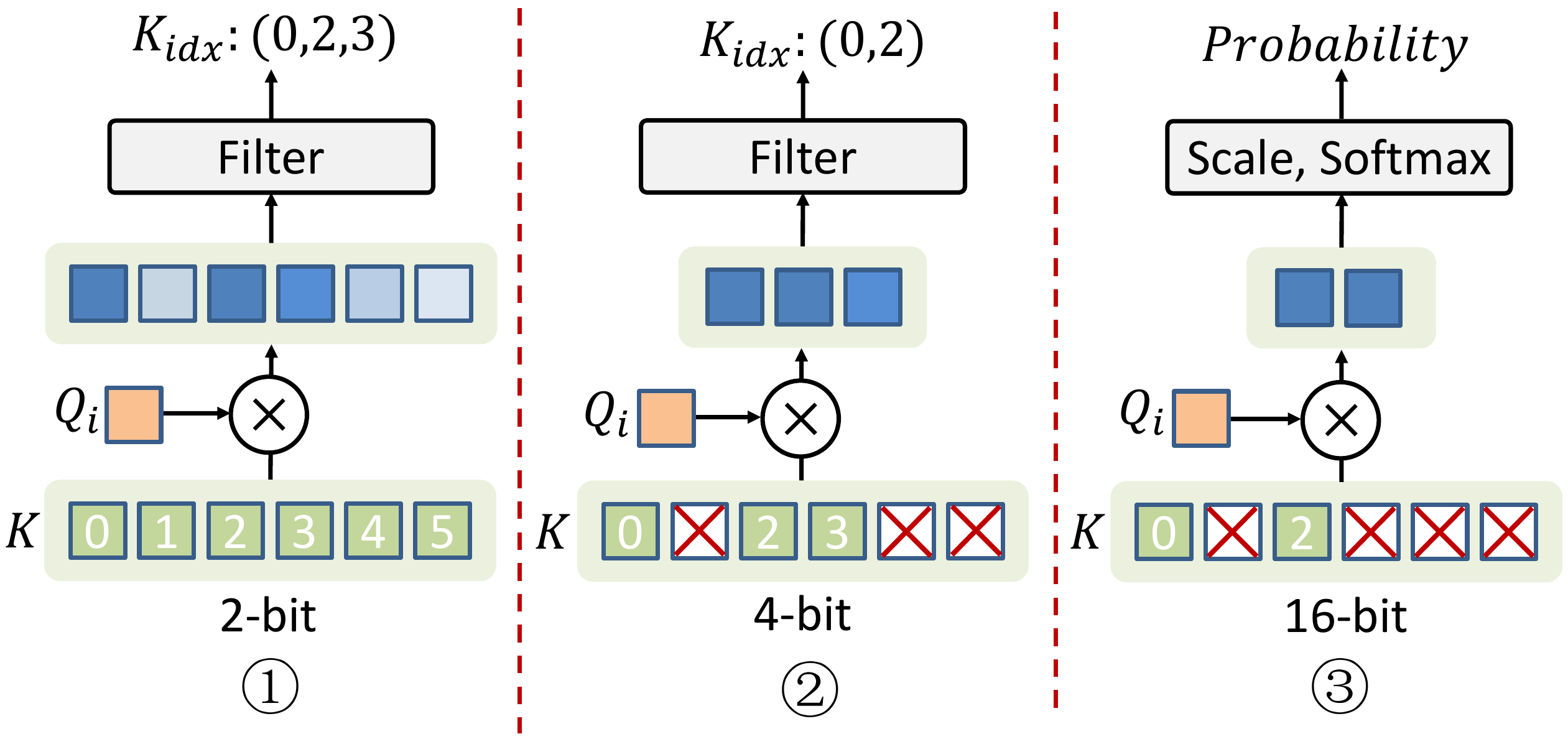} 
      \caption{An illustration of \trev{MP-MRF strategy.} \trev{Round {\textcircled{{1}}}: filter keys using 2-bit tensors.  Round \textcircled{{2}}: filter the selected keys further using 4-bit tensors.  Round \textcircled{{3}}: use the remaining keys to conduct high-precision sparse attention.  }}
            \label{fig:filtering}
\end{figure}

To overcome the aforementioned challenges, we propose a novel \trev{Mix-Precision} Multi-Round Filtering (\trev{MP-MRF}) method. As  Figure~\ref{fig:filtering} shows, for each query $Q_i$, we search for the important keys by setting multiple filtering rounds. In each round, we calculate the dot-production scores using low-precision tensors and select the keys whose scores are greater than a threshold. Only the finally selected keys are used for sparse attention with high-precision.  We introduce the details of  \trev{MP-MRF}
 mainly from the  three following aspects: 
 
(1) \textbf{Reduce Dot-production Overhead}: As pointed out above, the baseline top-k pruning still has to compute the whole $\emph{attention\_score}$. To mitigate the computational overhead, we can compute the dot-production $Q\cdot K^T$ with low-bit arithmetic.  However, directly applying low-bit quantization (e.g., ternary quantization) to the $Q,K$ tensors will significantly degrade the model accuracy because the attention features are sensitive to quantization errors~\cite{gobo, ternaryBert}.  To achieve a balance between efficiency and accuracy, we design a  \trev{Mix-Precision} Multi-Round Filtering (\trev{MP-}MRF) strategy. Instead of the one-round top-k pruning using high-precision tensors, we set multiple filtering rounds. In the beginning, we adopt extremely low bitwidth to compute $Q\cdot K^T$ and select the pairs with large scores. Although low-bit filtering is not that accurate, it still has good coverage if the pruning ratio is conservative (e.g., 50\%).  In the next round, we perform incremental filtering based on the selected pairs of the previous round. Since the remaining pairs tend to have closer scores, we use higher precision (more bits) to re-compute the remaining pairs' dot-production. We do such filtering iteratively. After the last round, the finally selected keys in each row are used to perform high-precision sparse attention. 

(2) \textbf{Avoid Top-k Selection}: 
Another problem is how to avoid the top-k selection. Instead of sorting all the scores, we can use mean-filtering~\cite{drq} to search for the important scores. Specifically, in each round, we estimate each row's mean value and only select the query-key pairs whose scores are greater than the mean value. In this way, we can roughly prune about 50\% of the elements in each round. Compared to top-k pruning, such a method is much simpler for implementation.

(3) \textbf{Adjustable Pruning Ratio}: 
The mean-filtering does not allow us to adjust the pruning ratio. 
However, in some scenarios, we expect high accuracy. In some other scenarios, we may slightly sacrifice accuracy for a higher speed. How to flexibly balance accuracy and pruning ratio according to different requirements is also an important problem.
Instead of directly adopting the mean values as thresholds, 
for round $r$, we set the threshold of the $i$-th row as $\theta_i^r$ :
\vspace{4pt}
\begin{equation}
\label{eq:thresh}
\myfont
    \theta_i^r  = \left\{\begin{aligned} \alpha_r\times max(S_i^r) + (1-\alpha_r)\times mean(S_i^r),~~~ 0\leq \alpha_r < 1  \\ 
    -\alpha_r\times min(S_i^r) + (1+\alpha_r)\times mean(S_i^r),  -1< \alpha_r < 0 
    \end{aligned} \right. 
\end{equation}
\vspace{1pt}

\noindent Where $S_i^r$ denotes the attention scores of the $i$-th row (the scores already pruned are ignored). We use a  parameter $\alpha_r$ to control the threshold,  whose value range is  ($-1, 1$). By adjusting  $\alpha_r$, the threshold transits from $min(S_i^r)$ to $max(S_i^r)$. Thus,  we can control the pruning ratio in each round from $0\%$ to $100\%$. We describe the overall \trev{MP-}MRF mechanism in Algorithm~\ref{algo:2}.

\begin{algorithm}[t]
\small

\label{algo:2}
\caption{\trev{Mix-Precision} Multi-Round Filtering }
\textbf{Input:} Total rounds $R$; \\ 
Filtering parameters \{$\alpha_0,...,\alpha_{R-1}$\};\\
Bit-width of each round \{$l_0,...,l_{R-1}$\};\\
Quantized $Q,K,V\in  \mathbb{R}^{n\times d}$; \\
\For{i $\leftarrow$~\emph{0} until $n$ }{
$K_{idx} \leftarrow \{0,1,..., n-1\}$; \\

\For{r $\leftarrow$~\emph{0} until $R$}{
\setstretch{1.1}
$Q_i' \leftarrow Q[i]$, \textcolor{blue}{\uline{load the first $l_r$ bits}};\\
$K'\leftarrow \text{gather}(K_{idx}, K)$, \textcolor{blue}{\uline{load the first $l_r$ bits}};\\
$S_i^r \leftarrow Q_i'\cdot K'^T$; \\  
Estimate $\theta_i^r$ using Equation \ref{eq:thresh}; \\
$K_{idx} \leftarrow \{j~|~S_i^r[j] > \theta_i^r\}$; }

\setstretch{1.0}
$Q_i \leftarrow Q[i]$, \textcolor{blue}{\uline{load full bits}};\\ 
$K' \leftarrow \text{gather}(K_{idx},K)$, \textcolor{blue}{\uline{load full bits}};\\
$V' \leftarrow \text{gather}(K_{idx},V)$;\\
\emph{attention\_prob} $\leftarrow$ Softmax$(Q_i\cdot K'^T/\sqrt{d})$\\
\emph{Result}[$i$] $\leftarrow $ \emph{attention\_prob}~$ \cdot ~V'$

}


\textbf{Return:} \emph{Result}
\end{algorithm}

(4) \textbf{Mix-precision quantization:} The proposed \trev{MP-}MRF requires mix-precision quantization. 
To make it more efficient, we first perform INT16 quantization  and then obtain the low-precision tensors (e.g., INT2 and INT4 tensors) directly by truncating the most significant bits (MSBs) of the INT16 data.  We only need to perform the quantization once and fetch a different number of bits to compute each filtering round. Moreover, such a way also enables us to reuse the results among different rounds, which will be discussed later.

We  illustrate a two-round \trev{MP-}MRF  in  Figure~\ref{fig:filtering}. Through experiments in Section~\ref{sec:round}, we observe that two-round filtering  with 2-bit tensors in the first round and 4-bit tensors in the second round is both accurate and cost-effective for our tasks. In the following discussions, we view the two-round filtering as our default configuration.

\section{Energon Accelerator}

\label{sec:accelerator}
\begin{figure*} [t]
    \centering
    \includegraphics[width=0.94\linewidth]{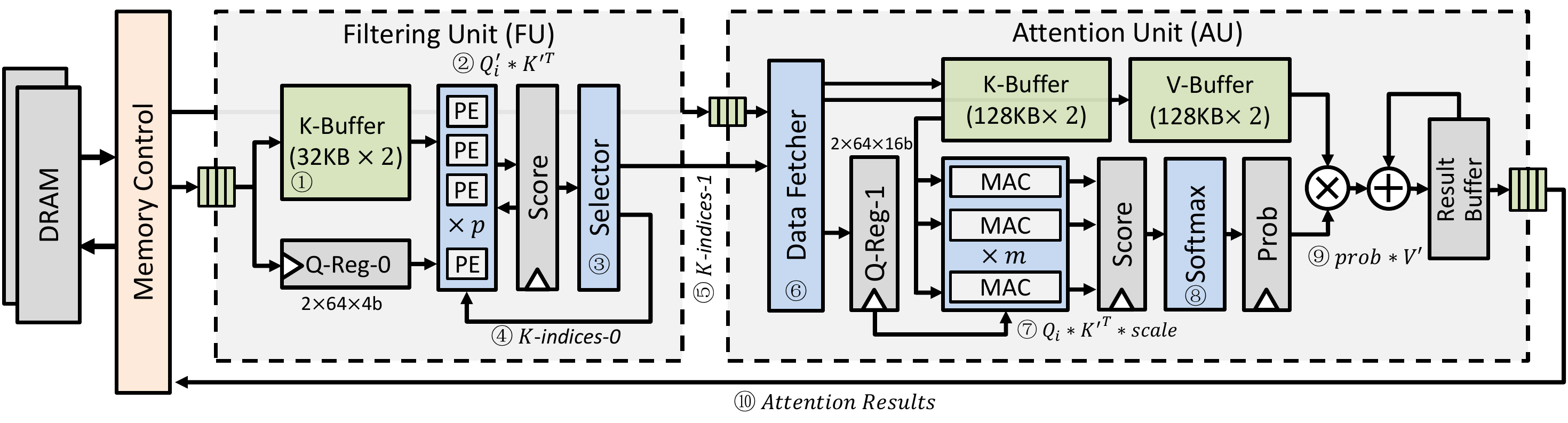} 
      \caption{Energon architecture overview.   \trev{The Filtering Unit (FU) performs multi-round mix-precision filtering to select important keys for each query. The indices of selected keys are fed into the Attention Unit (AU) for high-precision sparse attention. We adopt ping-pong buffers in both units to hide the memory access overhead as much as possible. }}
            \label{fig:overview}
            \vspace{-0.5em}
\end{figure*}

Current commodity  CPUs/GPUs and attention accelerators  can not support the proposed dynamic sparse attention mechanism  due to the mix-precision computation and specialized pipeline. Therefore, we propose a novel  Energon accelerator. It functions as a co-processor that is plug-in compatible with many other  NN accelerators.

Our Energon architecture mainly meets   two  design goals: (1) It supports the proposed MP-MRF with low latency and high throughput. We achieve this goal by designing efficient hardware pipelines and exploring data-level parallelism. (2) It is both energy-efficient and area-efficient. Since Energon is a co-processor working together with other \trev{NN accelerators}, it should save on-chip areas. We use specialized  IPU (Inner-Product Unit) with result-reusable PEs (\trev{Processing Elements}) to reduce arithmetic units' overhead to meet such goals.


\subsection{Architecture Overview}

\textbf{Design:}
 Figure~\ref{fig:overview} illustrates an overview of the Energon architecture.  In general, Energon is composed of two functional units. A Filtering Unit (FU) is designed for supporting the \trev{MP-}MRF mechanism (see Section~\ref{sec:filtering}) and outputs the indices of the important keys for each query.  The Attention Unit (AU) enables high-precision sparse attention, given the  $K$-indices (denotes the indices of selected keys).   Energon is equipped with external DRAM, which stores the $Q,K,V$ tensors  and the computed attention results. 

\textbf{Workflow:} 
\rev{When Q,K,V tensors are produced by other processors, they are also quantized and stored in DRAM, with INT4 and INT16 data stored separately. }Energon accelerator then computes one attention head at a time  and processes each head's queries in a pipelined manner. 
To be specific, when a head's processing starts, the $K,V$ tensors required by FU \rev{(INT4)} and AU \rev{(INT16)} are loaded from DRAM to on-chip buffers,  after which the queries are fed into the pipeline. Given a query $Q_i$, FU filters the important keys through cost-efficient low-precision operations and generates the indices of selected keys for AU.  AU performs high-precision sparse attention according to the received $K$-indices.  The attention result of each query will be written back to DRAM immediately.  Thus, co-located  DNN accelerators can use them to conduct the remaining computation. We introduce more details of each  module and the proposed optimizations in the following sections.


\subsection{Filtering Unit}

\begin{figure} [t]
    \centering
    \includegraphics[width=0.92\linewidth]{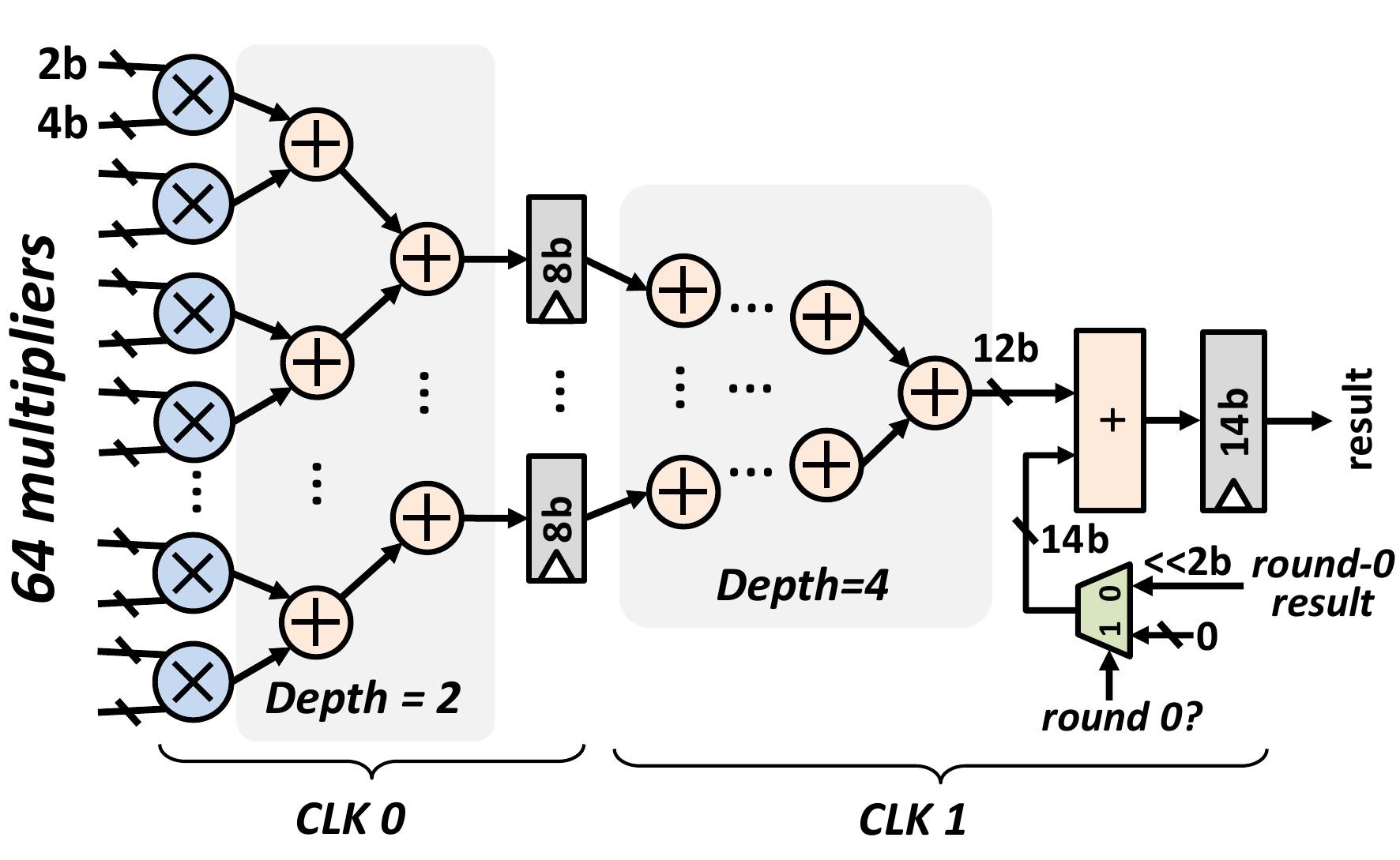} 
      \caption{\trev{ Architecture of the proposed Result-Reusable Multi-Precision PE.}}
            \label{fig:pe}
\end{figure}

The whole pipeline starts with the  Filtering Unit (FU). As aforementioned,  FU is designed to support the multi-round filtering algorithm. According to Algorithm~\ref{algo:2},  in the $r$-th round,   FU computes the approximated dot-production scores  $ S_i^r = Q_i'\cdot K'^T, i \in \{0,...,n-1\}$  using low-bit tensors (the adopted bitwidth in round $r$ is denoted  as $l_r$ ). It then searches for the indices of important keys by comparing them with a dynamic threshold $\theta_i^r$ estimated with \emph{min}, \emph{max} and \emph{mean} values (Equation~\ref{eq:thresh}). Under our settings, FU needs to perform two  filtering rounds in total. The first round is coarse-grained filtering using INT2 tensors. The selected keys are used for fine-grained filtering with INT4 tensors in the second round. Such a mix-precision workflow demands both INT2 and INT4 arithmetic units. To enhance the processing efficiency and save on-chip resources, we propose two specialized designs and optimizations:    result-reusable mix-precision IPU  and optimized key data layout. Details are introduced as follows: 

\textbf{IPU Design:} The IPU (Inner-Product Unit) marked as {{\textcircled{\small{2}}}} in Figure~\ref{fig:overview} is responsible for computing the inner-product $ Q_i'\cdot K'^T $ where $Q_i'$ is $1\times d$ vector  and $K'^T$ is $d\times n$ tensor (line 10 in Algorithm~\ref{algo:2}).  As we can see, IPU is composed of multiple  processing elements (PE) working in parallel. Each PE computes $d$ multiplications each time and adds up the results to get an inner-production result.   
To support both INT2 and INT4 processing, we design a novel PE architecture illustrated in Figure~\ref{fig:pe}. It not only enables mix-precision processing but also reduces the computational complexity through reusing the results. Specifically, supposing $d=64$ and a PE equips 64  multipliers. Each multiplier is responsible for a 4-bit$\times$2-bit multiplication. Instead of using 2-bit $Q$ in the first round (round-0) and 4-bit $Q$ in the second round (round-1), we adopt 4-bit $Q$ in both rounds while still keeping the $K$ to be 2-bit in the first round. Though such a change increases the computational overhead in round-0, it enables us to use the same hardware in both rounds and reuse the round-0 results.  

To be specific, given two 4-bit vectors $Q_i$ and $K^T_j$, we have  the following formulation: $$Q_i* K^T_j = (Q_i* K_j[3:2]<<2) + Q_i* K^T_j[1:0] $$ Where $K^T_j[3:2]$ means truncating the two most-significant bits (MSBs) of each element of $K_j$ while $K^T_j[1:0]$ represents the truncation of the two least-significant bits (LSBs).   Recall that we directly use the MSBs of INT4 data as the INT2 quantization. Therefore, if the INT4 operands ($Q_i$) are the same in both rounds, we can first calculate  $Q_i * K^T_j[3:2]$ in round-0, which serves as the round-0 score. Then in round-1, we calculate $Q_i*K^T_j[1:0]$, which is added with the left-shifted round-0 score to produce the round-1 score. By this means, we can save about half of the computation in round-1.  As shown in Figure~\ref{fig:pe}, we implement such a mechanism by equipping an adder and a shifter to deal with the shift-and-add operation and use a multiplexer to control the data path.  In round-0, both 2-bit and 4-bit operands are treated as signed integers, and the output of the adder tree will be added with zero. The results are  buffered in a Score register (see Figure~\ref{fig:overview}) and used to select important keys. In round-1, the 2-bit operands are viewed as unsigned integers. The corresponding round-0 scores are loaded, left-shifted, and added to the adder-tree results to obtain the final scores.

As shown in Figure~\ref{fig:pe}, we insert registers into the adder-tree to divide the whole PE into two stages. The multipliers and 2-level adder-tree belong to the first stage, while the 4-level adder tree and the shift-adder belong to the second stage. This design ensures that the latency of each stage is within $1 ns$ under the evaluation using the FreePDK 45nm standard library~\cite{freePDK}. Therefore, each PE outputs a result every two cycles. 

\textbf{Data layout:}  We store the 4-bit  keys in a $32$KB\trev{$\times2$ (double-buffering)} $K$-Buffer. 
Supposing that there are up to 1024 tokens ($n=1024$)  and $d = 64$ for each head, we store the MSBs and LSBs of each element separately. Each row of the SRAM has 512 bits containing 2-bit features from 4 tokens, which are transferred at a cycle.  IPU reads MSBs rows at round-0 and reads LSBs rows at round-1. To provide enough bandwidth in both rounds,
the MSBs and LSBs rows are interleaved to 8 single-ported SRAM banks. 
Moreover, the queries are stored in a register Q-Reg-0, the size of which is set to 2×64×4bits=64B. ‘2’ is for double buffering.

\begin{figure} [t]
    \centering
    \includegraphics[width=1.0\linewidth]{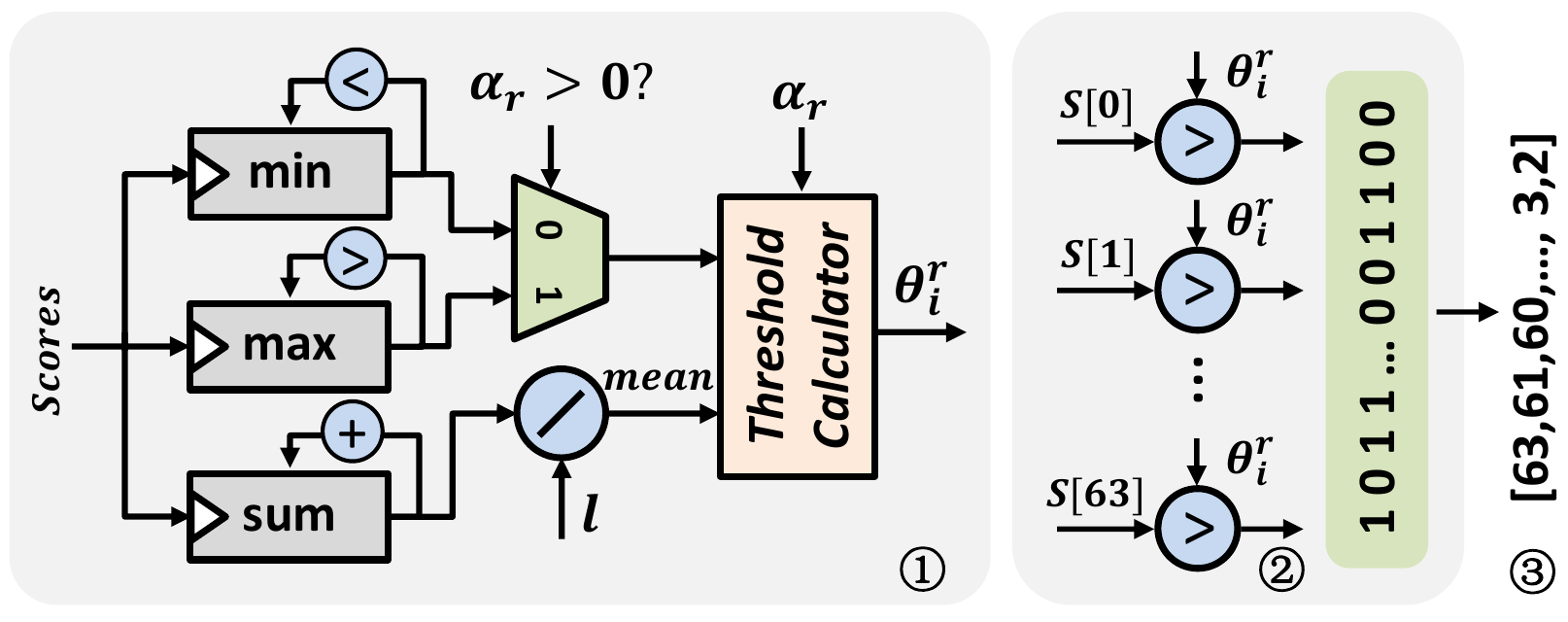} 
      \caption{\trev{Architecture of the Selector module.  \textcircled{{1}}: The \emph{min}, \emph{max} and \emph{sum} of scores are dynamically calculated to estimate the threshold $\theta_i^r$. \textcircled{{2}}: Scores are compared with the threshold in parallel to generate  the \emph{K-indices} \textcircled{{3}}. }}
            \label{fig:selector}
\end{figure}

\textbf{The Selector Module:}
\label{sec:selector}
As Figure \ref{fig:selector} illustrates, the Selector is responsible for selecting the important $K$-indices based on the dot-production scores computed by IPU.  It first estimates the dynamic threshold $\theta_i^r$ according to Equation \ref{eq:thresh}, then compares the  scores with the threshold.  Finally, the indices of keys whose corresponding scores are greater than $\theta_i^r$ are collected as the outputs.  Note that, Selector calculates \emph{min}, \emph{max} and \emph{sum} of the scores immediately after IPU outputs any results into the Score registers. Thus, it can finish estimating these three values right after all the dot-production scores of a query are calculated. The \emph{sum} value needs to be divided by $l$ (denotes the number of input scores) to get the \emph{mean} value. Then, it derives $\theta_i^r$ using a \emph{Threshold Calculator} logic, which conducts the calculation in Equation~\ref{eq:thresh} with the three estimated values and a filtering parameter $\alpha_r$.  After obtaining $\theta_i^r$, the scores are compared with it in parallel (the parallelism is 64 in the Figure). The output binary signals are converted to  \emph{K-indices}. As Figure~\ref{fig:overview} shows, the  \emph{K-indices-0} ({{\textcircled{\small{4}}}}) in the first round  are fed to IPU to conduct round-1 filtering, while the \emph{K-indices-1} ({{\textcircled{\small{5}}}}) are directly sent to AU  for high-precision sparse attention.

\subsection{Attention Unit}
As shown in  Figure~\ref{fig:overview}, the Attention Unit (AU)  relies on a Data-Fetcher to fetch the required keys and values from external DRAM according to the $K$-indices provided by FU, and stores them into the K-Buffer and V-Buffer, respectively.  Similar to FU, AU also equips a 2KB double-buffered Q-Reg to hold the 16-bit queries. A MAC Array with $m$ parallel MAC units calculates high-precision inner-productions. A Softmax module calculates the attention probabilities, whose exponential function is approximated using  Taylor expansion ~\cite{softmax}. Details about the whole attention pipeline and other optimizations are  introduced as follows:

\textbf{Attention Pipeline:} The attention operations are processed in a  pipelined manner. At each step, the query $Q_i$ and selected keys $K'$ are read from Q-Reg-1 and K-Buffer individually and sent to a MAC array to perform 16-bit scaled dot-production ($Q_i\cdot K'^T\cdot scale$, where $scale = \frac{1}{\sqrt{d}}$). The results are buffered in the Score registers. A Softmax module ({{\textcircled{\small{8}}}}) then processes the attention scores to get the attention probabilities. Each Softmax module consists of 8  exponential units working in parallel to improve throughput. After getting probabilities (Prob), AU finally computes attention results by multiplying Prob and the selected values $V'$ through another multiply-accumulate module ({{\textcircled{\small{9}}}}). The  attention results  ({{\textcircled{\scriptsize{10}}}}) are  written back to DRAM. Since we adopt double buffers to overlap the loading and computation of queries, the whole workflow is fully pipelined when processing an attention head. 
Unlike SpAtten, we do not double $K,V$ buffers since the loading of $K,V$ and computation are always imbalanced for our long-input cases. Details about a performance model will be discussed in Section~\ref{sec:analytic}.  


\textbf{On-Demand Fetching:} Recall that $A^3$ needs to load all keys and values on-chip for approximated attention, which involves tremendous DRAM access. However,  
we notice that after multi-round filtering,  if the pruning ratio is relatively high,  more than half of the keys in each head are "always unimportant" that don't need to be loaded. For instance, when applying 16$\times$ pruning, BERT-base on  Squad-v1  dataset only needs averagely 47\% of keys for attention in each head.  Thus, instead of always fetching all the $K,V$ features, we propose an On-Demand Fetching (ODF) strategy, which only fetches the keys and values needed by each query incrementally.

For each head, we first load the keys and values required by the first query according to the $K$-indices provided by FU. After one query is processed, the Data Fetcher module analyzes the required  $K$-indices of the following query and checks whether any keys and values are not in the buffer. If so, only the missed keys and values will be loaded. 
Moreover, the keys already in $K$-Buffer will be computed immediately to overlap DRAM access  with computation for each query.

\begin{figure} [t]
    \centering
    \includegraphics[width=1.0\linewidth]{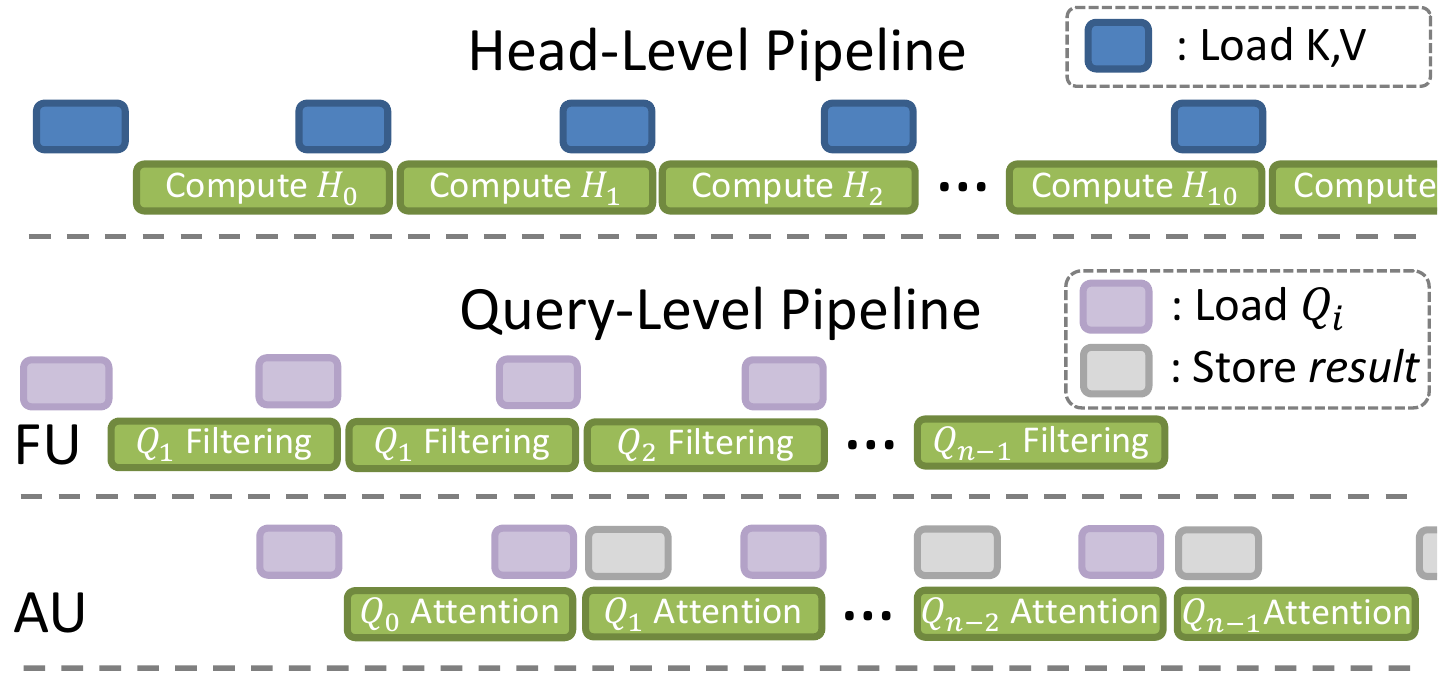} 
      \caption{Two-Level Pipelines. \trev{Double QKV buffers enable the overlapping of memory access and computation.}}
            \label{fig:pipeline}
\end{figure}

\subsection{Performance Model}
\label{sec:analytic}
According to Energon's workflow, there are two processing pipelines: the head-level pipeline and the query-level pipeline.  As illustrated in Figure~\ref{fig:pipeline}, each head loads the required $K,V$ tensors on-chip and then starts computing. After finishing one head,  the next head will load the $K,V$ and continue the computation. We call it head-level pipeline.  Within each head, the queries' loading and the computation are overlapped for both FU and AU, since double Q-Regs are adopted. The filtering and attention for each query are also pipelined between FU and AU. We call such a workflow query-level pipeline.  

 \begin{figure*} [t]
    \centering
    \includegraphics[width=1.0\linewidth]{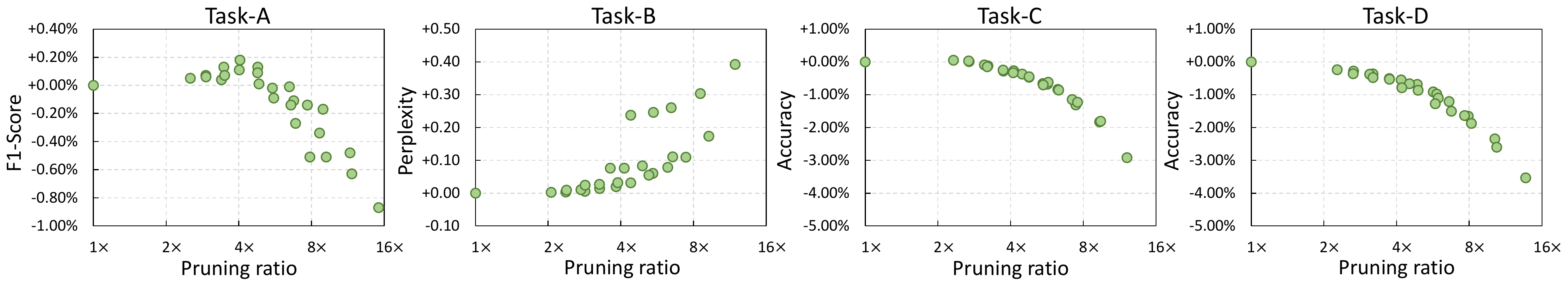} 
    \vspace{-1em}
      \caption{Accuracy  VS. pruning ratio exploration using the proposed \trev{MP-MRF algorithm. }} 
            \label{fig:accuracy}
             \vspace{-1em}
\end{figure*}

For different workloads, the sequence lengths, embedding dimensions, and pruning ratios vary a lot. Then what will be the bottleneck of the pipeline? We propose a performance model to answer this question. Assuming that the DRAM bandwidth is $B$ bytes per cycle. Loading $K,V$ data of each head costs about  $t\_load = \frac{4.5\times d \times n}{B}$  cycles, where 4.5 is derived by loading 4 bytes for both $K,V$ elements in AU and  0.5 byte for each $K$ element in FU. Supposing the pruning ratio is $\beta$, the fully pipelined  AU processes each query using $\frac{2\times \beta \times  n}{m}$ cycles as the MAC array outputs $m$ results every two cycles. Thus, the computation of each head costs $t\_comp = \frac{2\times \beta \times  n\times l}{m}$ cycles, where $l$ is the query length. For text-generation tasks with cached $K,V$ (enable \texttt{use\_cache} in the hugging-face implementation), $l$=1, otherwise $l$=$n$.  Thus, the loading-to-computation ratio is  roughly $\frac{t\_load}{t\_comp} = \frac{2.25\times d\times m}{B\times \beta \times l}$. \trev{For some typical values, say  $d=64$, $m=8$, $l=512$ and $\beta=0.25$, if the HBM memory  with 512GB/s  bandwidth (similar to the setup in SpAtten) is equipped and the functional units runs at 1GHz,  then the loading-to-computation ratio is merely $0.017$. If low-bandwidth DRAM like LP-DDR3 is equipped, we assume the bandwidth is 25.6GB/s (two channels), then the ratio is just $0.35$}. \trev{That is to say, for long input sequences,  the loading time is usually much less than the computation time. Thus the head-level pipeline is computation-bounded. Double buffers can hardly bring too much performance gain. We can therefore disable half of the double K-V buffers through clock-gating to save energy. For middle and short input sequences, say $l=128$, the $t\_load : t\_compute$  ratio becomes 1.44 (with LP-DDR3 memory). Considering that the data loading and computation cycles are of the same order of magnitude, we still enable double-buffering to maximum  throughput. Therefore, we can always use the  $t\_load : t\_compute$ ratio to decide whether to enable double-buffering under different configurations and workloads to either save energy or optimize throughput.} 

To balance the query-level pipeline,  the filtering cycles should match the attention cycles for each query. Since the IPU parallelism is $p$,  it computes each $Q_i$ in $t\_comp' =  \frac{2\times (1+\gamma)\times  n }{p}$ cycles, where $\gamma$ denotes the pruning ratio in round-0. Considering that the Selector's cycles are negligible if it is highly parallelized,  we directly use $t\_comp'$ to estimate the filtering cycles for each query. We let  $t\_comp'$ =  $t\_comp$ for balancing the FU-AU pipeline, then we derive $\frac{m}{p} = \frac{\beta}{1+\gamma}$. Since  $\frac{\beta}{1+\gamma}$ is always much smaller than 1 according to our experiments, FU should have higher parallelism than AU to balance the whole pipeline. Such a goal is easy to achieve since FU only involves cost-efficient low-precision arithmetic units, which are easy to scale up. \rev{We leverage such a performance model to determine hardware configurations in Section~\ref{sec:arc_eval}}.  





\section{Evaluation}
\label{sec:eval}




\subsection{Algorithm Evaluation}

\label{sec:eval1} 


\textbf{Benchmarks:}
To demonstrate the efficiency of the proposed dynamic sparse attention algorithm, we evaluate it on both NLP and CV tasks.  We prepare four representative benchmarks listed in Table~\ref{tab:benchmark}. The aforementioned NLP tasks, namely the question-answering task on SQuAD-v1~\cite{squad} dataset and the language-modeling task on Wikitext-2~\cite{wikitext} dataset serve as tasks A and B, respectively.  We also adopt two image-classification datasets: CIFAR-100~\cite{cifar} and ImageNet~\cite{imagenet}. The input sequence lengths are also listed in the table.  SQuAD has variable sequence lengths among  samples, and the 95th percentile of sequence length is 304 (average length is 176). Task-B with Wikitext-2 dataset has 1024 tokens, and the two CV datasets all have 577 input tokens. We do not adopt GLUE dataset~\cite{GLUE} for evaluation, because each GLUE benchmark task only has a limited sequence length ($n<$  100). Obviously, for such tasks, the attention operations will not be the bottleneck. 
\begin{table}[t]
\vspace{-0.5em}
\caption{Benchmarks}
\renewcommand\arraystretch{1.2}
\label{tab:benchmark}
\resizebox{0.48\textwidth}{!}
{
\begin{tabular}{|ccccc|}
\hline
\textbf{Task} &\textbf{Type}  & \textbf{Dataset}  & \textbf{Model} & \textbf{Sequence Length}\\ 
\hline 
A  & NLP    & SQuAD-v1~\cite{squad}     & BERT-base~\cite{bert}     &   304 (95th pctl.)  \\ 
B  & NLP    & Wikitext-2~\cite{wikitext}   & GPT-2~\cite{gpt2}         &   1024              \\ \hline
C  & CV     & CIFAR-100~\cite{cifar}    & ViT-B/16~\cite{vit}     &   577               \\ 
D  & CV     & ImageNet~\cite{imagenet}     & ViT-L/16~\cite{vit}     &   577          \\ \hline
\end{tabular}
}
\centering
\end{table}

\textbf{Model Setup:} We use BERT-base for Task-A, GPT-2  for Task-B and adopt ViT-B/16 and ViT-L/16 ~\cite{vit} for  Task-C and Task-D. All the models except for ViT-L/16 have 12 transformer blocks, each having 12 attention heads. The ViT-L/16 model has 24 blocks with 16 heads in each block. For all the models,  each head's feature dimension $d$ is 64. We implement the two NLP models using the hugging-face transformers project, while implement the ViT models using the Pytorch-ViT project~\cite{pytorch_vit}. The input image resolution of Task-C and Task-D are resized to $384\times 384$. They are chunked into 576  patches and prefixed with a \texttt{Start} token to serve as input sequence.  All the models are trained on the officially released pre-train weights with default training parameters. We make sure that the trained models have the same accuracy as the claimed performance in these projects. We then modify the code to support the proposed dynamic sparse attention. 


\textbf{Accuracy and Pruning Ratio Exploration:}
We first evaluate the accuracy and pruning ratio by exploring different parameters $\alpha_r$. For each round we set  $\alpha_r$ from -0.2 to 0.2 with a  step of 0.1. Thus, there are five different parameters per round and 25 configurations in total.  We estimate the pruning ratio and accuracy of each configuration. Since this exploration only involves inference on test sets, it takes several minutes to hours to finish. We plot all the exploration results in Figure~\ref{fig:accuracy}. For each scatter plot, the X-axis represents the pruning ratio, and the Y-axis is the accuracy (For Task-B, the Y-axis represents the perplexity, lower is better).  The optimal configuration should have both a high pruning ratio and accuracy. 

With the exploration results, we have the following observations: 
(1) For NLP tasks, namely Task-A and Task-B, we can get up to 11.5$\times$ and 9.25$\times$ pruning ratio with negligible accuracy loss (-0.48\% F1-Score for Task-A and +0.17 perplexity for Task-B). On the two CV tasks, we also obtain $4.8\times$ and $3.7\times$ pruning with less than $0.5\%$ accuracy loss. Moreover, even if we do not tolerate any accuracy loss, our algorithm still provides more than $2\times$ pruning except for Task-D.  Generally, the NLP tasks achieve higher pruning ratios than the CV tasks. This may be because natural language contains more redundancy than images in these tasks.
(2) For Task-A, $4\times$ pruning even improves the overall accuracy. We argue that pruning helps the model be more concentrated on those important query-key pairs, which is  meaningful for question-answering tasks.  (3)  CV tasks are less sensitive to the pruning configurations. As we can see, Task-C and Task-D's distributions are more smooth than that of Task-A and Task-B. We argue that there are fewer "dominant tokens" for the image-classification tasks that, if wrongly pruned, will significantly degrade the accuracy  (e.g., the tokens containing the question-answer for QA tasks).  
We choose the configuration with the highest pruning ratio and negligible accuracy loss (within 0.5\%) as our best configuration for each task and conduct the following  evaluations. 

\begin{table}[t]
\caption{\rev{Top-k Coverage Analysis}}
\vspace{-4pt}
\label{tab:coverage}
\small
\resizebox{0.48\textwidth}{!}
{
\begin{tabular}{|lllll|}
\hline
Tasks                 & Task-A       & Task-B       & Task-C       & Task-D       \\ \hline
Optimal pruning ratio & 11.5$\times$ & 9.25$\times$ & 4.77$\times$ & 3.73$\times$ \\ 
Top-k coverage ratio        & 97.3\%         & 91.1\%        & 95.1\%        & 96.0\%        \\ \hline
\end{tabular}}
\vspace{-4pt}
\end{table}

\label{sec:coverage}
\rev{
\textbf{Top-k Coverage Analysis:} To better understand the effectiveness of our \trev{MP-}MRF strategy, we also evaluate the overlap between the set of Q-K pairs selected by \trev{MP-}MRF and true top-k pairs obtained by exhaustive search. As shown in Table~\ref{tab:coverage}, we observe that under the optimal  pruning ratio, the top-k coverage is higher than 95\% for Task-A, Task-C and Task-D. Even for Task-B, which has the longest sequence length, the coverage ratio reaches about 91.1\%. Therefore, our \trev{MP-}MRF serves as a practical approximation of top-k selection. }

\subsection{Architecture Evaluation} 
\label{sec:arc_eval}

\begin{table}[]

\caption{Hardware and Architecture Parameters}
\vspace{0pt}
\renewcommand\arraystretch{1.25}
\label{tab:parameter}
\resizebox{0.48\textwidth}{!}
{
\begin{tabular}{|c|c|}
\hline
\textbf{Module} & \textbf{Configurations}                                                                                                           \\ \hline
\multicolumn{2}{|c|}{\textbf{Energon-edge}}                                                                                                         \\ \hline
DRAM            &2-Channel LP-DDR3-1600, Bandwidth = 25.6GB/s;                                                                                     \\ \hline
Filtering Unit  & 32KB \trev{$\times 2$}  K-Buffer; 8-PE IPU;                                                                                                           \\ \hline
Attention Unit  & \begin{tabular}[c]{@{}c@{}}128KB \trev{$\times 2$}   K/V-Buffer; 1$\times$ MAC;\\ 1$\times$ Softmax; 64 Multipliers for $prob\times V$;\end{tabular}    \\ \hline
\multicolumn{2}{|c|}{\textbf{Energon-server}}                                                                                                       \\ \hline
DRAM            & HBM-1.0, Bandwidth =  256 GB/s;                                                                                                                     \\ \hline
Filtering Unit  & 32KB \trev{$\times 2$}  K-Buffers; 64-PE IPU;                                                                                                          \\ \hline
Attention Unit  & \begin{tabular}[c]{@{}c@{}}128KB \trev{$\times 2$}  K/V-Buffers; 8$\times$ MAC;\\ 8 $\times$ Softmax; 512 Multipliers for $prob\times V$;\end{tabular} \\ \hline
\end{tabular}
}
\centering
\end{table}

\textbf{Methodology:} We evaluate two configurations namely  Energon-edge and Energon-server to meet the requirements in edge-computing and cloud-computing scenarios. The hardware and architecture parameters are listed in Table~\ref{tab:parameter}.  Energon-edge has much fewer arithmetic units compared to Energon-server and equips LP-DDR3 DRAM to reduce the power. According to our performance model in Section~\ref{sec:analytic}, the ratio of AU's parallelism $m$ to FU's parallelism $p$ should better be close to $\frac{\beta}{1+\gamma}$, where $\gamma$ and $\beta$ are the pruning ratios of round-0 and round-1. Our algorithm evaluation reveals that $m:p$ = $1:8$ is the most suitable for each tested task. \trev{Also, according to the estimated $t\_load : t\_compute$ ratios, we enable double-buffering on Task-A and disable half of the buffers on the remaining tasks by clock-gating.}

We implement Energon with Chisel~\cite{chisel} and compile it to Verilog RTL. We further synthesize the RLT  using Synopsys Design Compiler under FreePDK 45nm standard library ~\cite{freePDK} to estimate the logic parts' area and power consumption. The power, area and read/write bandwidth of on-chip  SRAM buffers are estimated through CACTI~\cite{cacti}. For modeling off-chip DRAM, we simulate the memory behaviors with Ramulator~\cite{ramulator}.
According to the synthesized results, the latency of the critical path is less than 1$ns$. Then we assume the running frequency of Energon is 1GHz.  We extract each stage's actual cycles by simulating the RTL with Verilator~\cite{snyder2004verilator}, based on which a cycle-level simulator is implemented to evaluate end-to-end performance. 

We compare Energon with several general-purpose computation platforms. We adopt Intel Xeon  5220  CPU~\cite{XEON} and NVIDIA Tesla-V100 GPU~\cite{V100} as baselines to evaluate Energon-server. To evaluate Energon architecture's efficiency against mobile platforms, we also compare Energon-edge with  NVIDIA Jetson-TX2 GPU~\cite{TX2} and ARM A72 CPU (4 cores, Raspberry Pi 4 Modle B platform~\cite{raspberry}). We deploy the benchmarks on these platforms using  Pytorch framework~\cite{pytorch} and record the execution time of attention operations through inserting \texttt{torch.cuda.synchronize()} at the start and end points of attention layers and then calculate the spent time. \rev{We conduct the evaluation using Pytorch-1.7 with   CUDA-10.2 (cudnn 7.6) and MKL-2020.2 as backends. } \rev{Note that to test the latency of generation model GPT-2 (Task-B), we follow SpAtten's practice and set the initial length  as 992 and measure the latency of generating 32 tokens.}

\begin{figure} [t]
    \centering
    \includegraphics[width=1.0\linewidth]{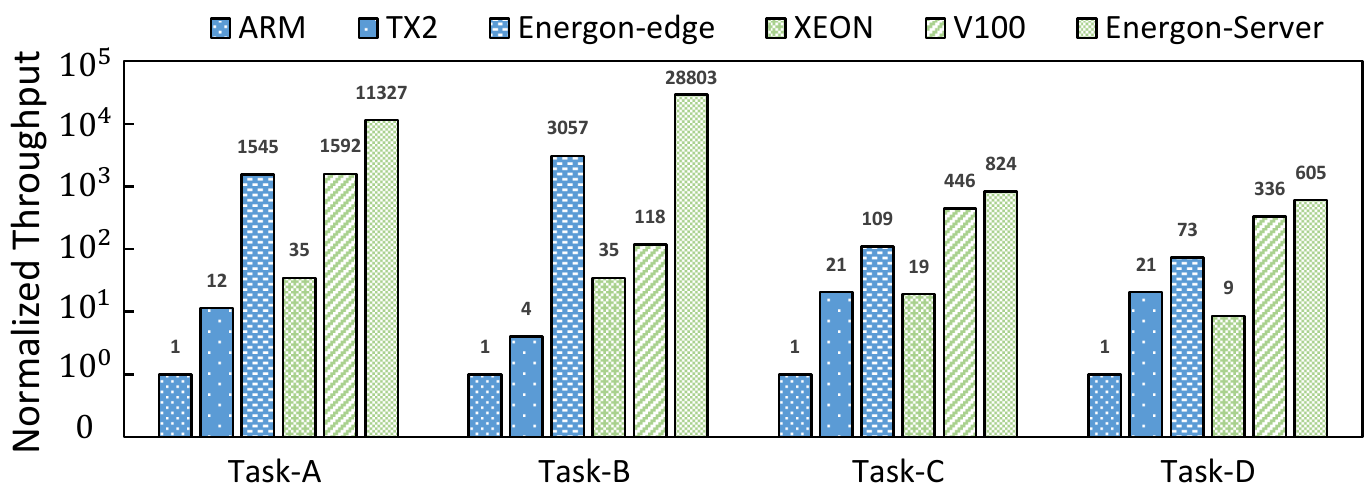} 
      \caption{\trev{Normalized throughput comparison.}}
            \label{fig:speedup}
\end{figure}

\begin{figure} [t]
    \centering
    \includegraphics[width=1.0\linewidth]{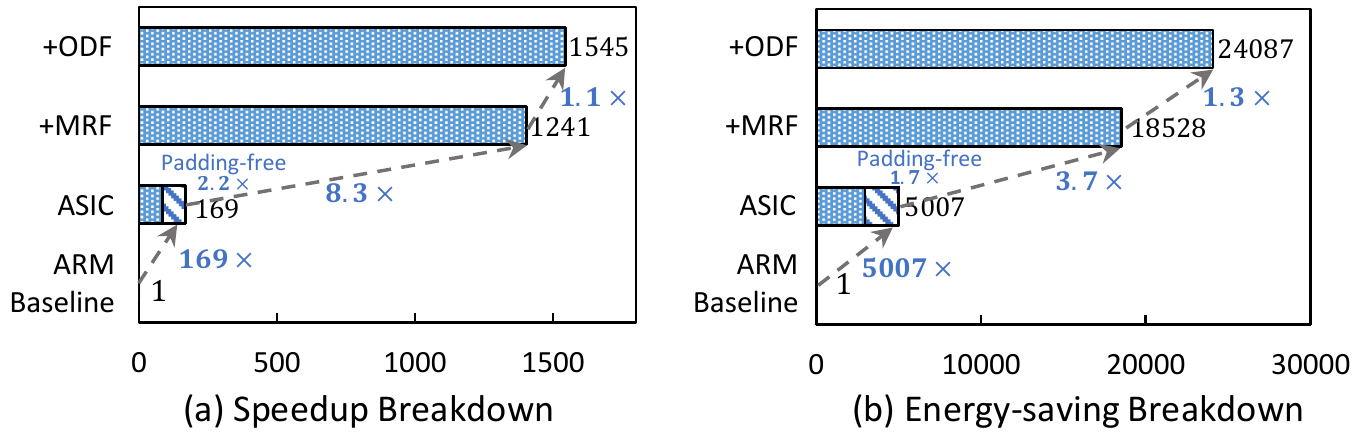} 
      \caption{\rev{Speedup and Energy-saving Breakdown.} }
            \label{fig:breakdown}
\end{figure}

\textbf{Throughput Improvement:}
We estimate the throughput of Energon accelerators on all the benchmarks via simulation. The speedup over baseline platforms is shown in Figure~\ref{fig:speedup}. 
\rev{ As we can see,  Energon-edge achieves 73$\times$ \trev{(Task-D)} to $3057\times$ \trev{(Task-B)} speedup over ARM A72 CPU among the four tasks and achieves  \trev{3.4}$\times$\trev{(Task-D) to} $\trev{764\times}$\trev{(Task-B)}  speedup against TX2 GPU. Compared to Xeon CPU and V100 GPU, Energon-server shows \ServerGeomeanSpeedupOverXeon~ and \ServerGeomeanSpeedupOverGPU~ geo-mean speedup. Our proposed Energon architecture obtains considerable throughput improvement.}
\rev{
Figure\ref{fig:breakdown}-a shows the speedup
breakdown of Energon-edge over  ARM CPU on Task-A. As we can see, the specialized hardware pipeline is  \trev{$169\times$} faster than ARM CPU. The speedup partially comes from our padding-free dataflow ($2.2\times$).  The CPU baseline running hugging-face framework cannot deal  with  SQuAD dataset well, whose input lengths vary a lot among samples and thus requires padding. Our \trev{MP-}MRF and ODF strategies bring $8.3\times$ and $1.1\times$  speedup, respectively. Considering that task-A is a computation-intensive task, \trev{MP-}MRF successfully reduces overall latency. 
}

\textbf{Area, Power, and Energy:}
We estimate the area and power of Energon  and list the breakdown of the core parts in Table~\ref{tab:ppa}. The area and power are reported by Design Compiler using FreePDK 45nm  library~\cite{freePDK}. The overhead of on-chip SRAMs is estimated using CACTI~\cite{cacti}.  As we can see, Energon-edge and Energon-server utilize about \trev{4.20} mm$^2$ and \trev{8.62} $mm^2$ areas, respectively.
For Energon-server and Energon-edge, the Filtering Units only occupy \trev{$10.41\%$} and \trev{$31.64\%$} of  total area, and consume \trev{$12.39\%$} and \trev{$29.23\%$} of  total power, which demonstrates the efficiency of the proposed multi-round filtering mechanism. Adopting low-bitwidth arithmetic in FU and the result-reusable mix-precision PEs jointly save the power and area.
\rev{In Table \ref{tab:power_breakdown} we also list the peak power consumed by the memory interface and external DRAM}.

\begin{table}[t]
\caption{ Area and Power Breakdown. \rev{(Core Part)} }
\vspace{-4pt}
\small
\renewcommand\arraystretch{1.2}
\label{tab:ppa}
\resizebox{0.48\textwidth}{!}
{
\begin{tabular}{|cccc|}
\hline
\textbf{Module}                          & \textbf{Component}   & \textbf{Area(\%)} & \textbf{Power\%} \\ \hline 
\multirow{3}{*}{Filtering Unit} & Arithmetic &     \trev{2.96} $|$ \trev{2.22}$^*$         &      \trev{4.28} $|$ \trev{6.55}$^*$     \\ 
                                & Buffer      & \trev{7.31} $|$ \trev{29.13}$^*$       &   \trev{7.95} $|$ \trev{22.23}$^*$        \\
                                & Control     & \trev{0.14} $|$ \trev{0.29}$^*$       & \trev{0.16} $|$ \trev{0.45}$^*$      \\ \hline
\multirow{3}{*}{Attention Unit} & Arithmetic &     \trev{27.14} $|$ \trev{9.31}$^*$         &    \trev{30.88} $|$ \trev{10.79}$^*$      \\
                                & Buffer      &   \trev{62.21} $|$ \trev{58.55}$^*$          &    \trev{56.44} $|$ \trev{59.16}$^*$ \\ 
                                & Control     &      \trev{0.25} $|$ \trev{0.50}$^*$       &       \trev{0.29} $|$ \trev{0.81}$^*$    \\ \hline

\multicolumn{2}{|c}{Total}                   &     \trev{8.62} $|$ \trev{4.20}$^*$ (mm$^2$)        &   \trev{0.89} $|$ \trev{0.32}$^*$ (W)       \\ \hline
\end{tabular}
}
$~~~~~~^*$ Energon-edge
\vspace{-1.5em}
\end{table}

\begin{table}[t]
\caption{\rev{Power Breakdown of Energon}}
\label{tab:power_breakdown}
\small
\resizebox{0.48\textwidth}{!}
{
\begin{tabular}{|lcccc|}
\hline
Tasks                 & Core Part      &  Memory Interface  & DRAM      & Total       \\ \hline
Energon Server &  \trev{0.89W}  & 2.4W\cite{hbminterface} & 7.3W\cite{hbm} & \trev{10.6W} \\ 
Energon Edge   &  \trev{0.32W}  &  0.9W\cite{lpddr3Controller}  &  1.5W\cite{LPDDR3} &    \trev{2.7W}     \\ \hline

\end{tabular}}
\end{table}

\begin{figure} [t]
    \centering
    \includegraphics[width=1.0\linewidth]{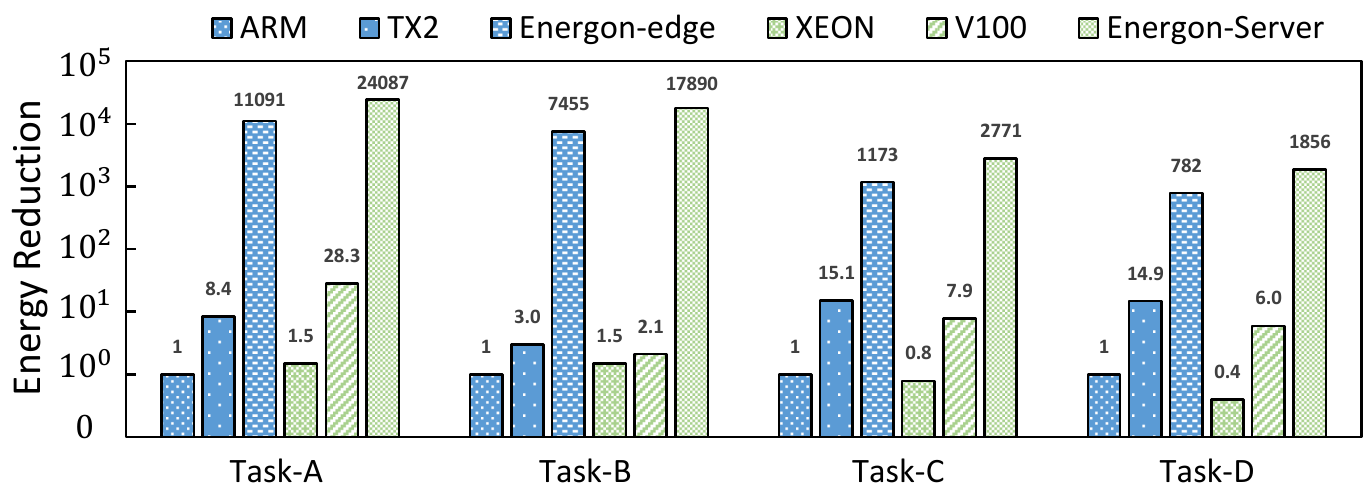}
    \vspace{-1.5em}
      \caption{\rev{Energy Saving Over Baselines.}}
            \label{fig:energy_saving}
             \vspace{-0.5em}
\end{figure}

We further evaluate the end-to-end energy consumption of Energon accelerators and the baseline platforms on our benchmarks. \rev{The energy of the core parts is estimated using \emph{latency} $\times $ \emph{power}, while the energy consumed by the  memory interface and DRAM is estimated according to total memory access. Specifically, we derive the energy consumption of DRAM using Ramulator and DRAMPower~\cite{chandrasekar2012drampower}.   We profile the energy consumption of Xeon CPU with PyRAPL~\cite{PyRAPL}. The running power of TX2 GPU and NVIDIA V100 GPU are estimated using PyNVML~\cite{PyNVML}.  For ARM A-72@4-Cores, we estimate the power to 4W.}
We plot the energy reductions on four tasks in Figure~\ref{fig:energy_saving}. Thanks to their high throughput and low power usage, Energon-edge and Energon-server are   more energy efficient by orders of magnitude ($10^3$-$10^4\times$) compared to the other general-purpose platforms. \rev{In Figure~\ref{fig:breakdown}-b we also show the energy-saving breakdown. \trev{MP}-MRF and ODF bring $3.7\times$ and $1.3\times$ energy-saving, respectively. We also find that on memory-intensive tasks like Task-B, ODF  saves more energy (up to $1.5\times$), which is not shown in the figure.}

\subsection{Comparisons with Existing Designs}

\label{sec:comparison}

 \begin{figure} [t]
    \centering
    \includegraphics[width=1.0\linewidth]{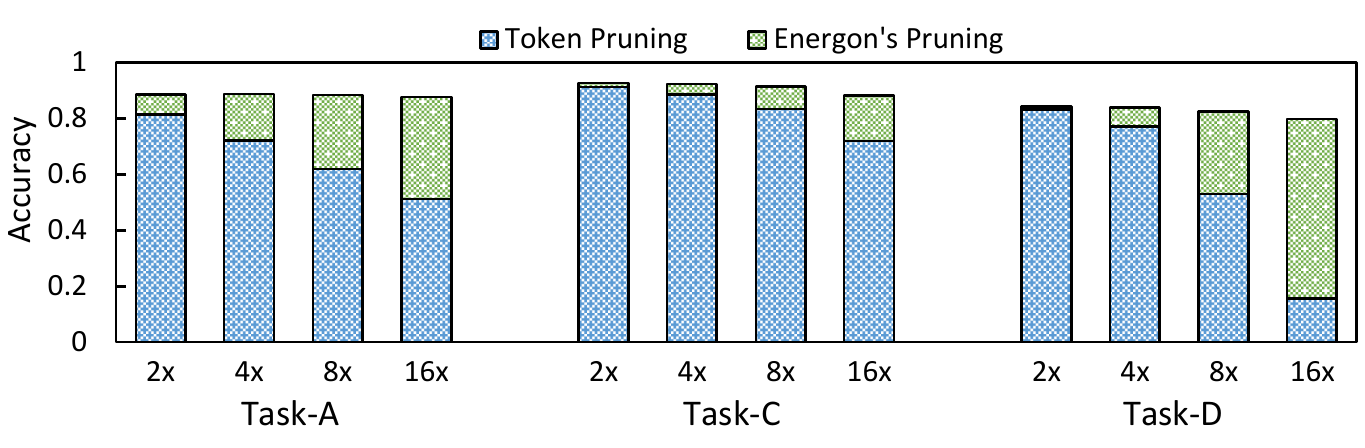} 
    \vspace{-1.5em}
      \caption{\rev{Accuracy Comparison with \trev{SpAtten's} Token-Pruning.}}
            \label{fig:compare_Spatten}
                  \vspace{-0.5em}
\end{figure}

\textbf{Comparison with SpAtten:}
SpAtten~\cite{spatten} 
adopts the \trev{cascaded}
token-pruning to remove "unimportant" tokens permanently. It accumulates attention probabilities among layers and prunes the tokens with low accumulative probabilities.
However,  such a coarse-grained pruning method suffers significant accuracy loss without retraining.  
\rev{We compare the \trev{MP-}MRF method with token-pruning on our benchmarks.} For a fair comparison, we implement both methods with the same framework and adjust the pruning parameters to ensure that they have the same pruning ratio.  We do not prune the rows of attention matrices when implementing token pruning, which otherwise will result in extremely low accuracy without retraining.

\rev{As we can see in Figure~\ref{fig:compare_Spatten}, SpAtten shows lower accuracy on the three accuracy-oriented tasks. When the pruning ratio reaches $16 \times$, our method's accuracy is on-average 2.32$\times$  higher than token-pruning's.  With the same accuracy, Energon achieves averagely $5.3\times$ higher  pruning ratio. We assume SpAtten has $2\times$ pruning ratio on the three tasks and Energon has $16\times, 8\times$, $8\times$ pruning ratios on Task-A,C,D. We  assume  they all equip HBM-1.0 and the MACs have the same computational capacity. Since SpAtten's head-pruning and progressive quantization also contribute $1.1\times$ and $2.8\times$ speedup,  thus  Energon-server should have $5.3/(1.1\times2.8)$ = $1.7\times$ higher effective throughput over SpAtten.  Moreover, Energon has $1.6\times$ higher energy efficiency, assuming SpAtten is {9.9w} (7.3W HBM-1.0 + 2.6W core \cite{spatten}).}

\textbf{Comparison with $A^3$:} 
$A^3$ ~\cite{a3}  adopts several numerical approximations to reduce the overall complexity. However,
$A^3$ needs to load all the data from DRAM to perform candidate selection and post-scoring selection. It consumes more energy than Energon, especially when the pruning ratio is high. For example, under the \emph{aggressive} mode of $A^3$, it gets about $22\times$ pruning ratio on the SQuAD-v1 dataset (Bert-base model). We adjust the pruning parameters of Energon and evaluate the accuracy and DRAM access reduction with the On-Demand Fetching strategy. Since $A^3$ adopts 8-bit quantization, to make the comparison fair, we also adjust the attention bit-width to 8-bit.   Our Energon accelerator achieves a sightly higher F1-Score under the same pruning ratio ($0.869$ VS. $0.867$). Thanks to the ODF strategy, Energon saves about $34.9\%$ of total DRAM read and achieves 1.35$\times$ total DRAM access reduction. Assuming that both accelerators equip 2-channel  LPDDR3-1600 DRAM,  Energon saves about 19.8\% of total DRAM access energy. Moreover, we downsize Energon-edge to make it have the same amount of $K,V$ buffer with $A^3$. According to our estimation, Energon saves about $1.5\times$ energy and requires $1.2\times $ smaller area  compared to $A^3$. \rev{Assume A$^3$ and Energon have the same pruning ratio and computation ability, Energon also shows  $1.25\times$ average speedup against $A^3$, due to the saving of data-loading time.} 

\begin{figure} [t]
    \centering
    \includegraphics[width=1.0\linewidth]{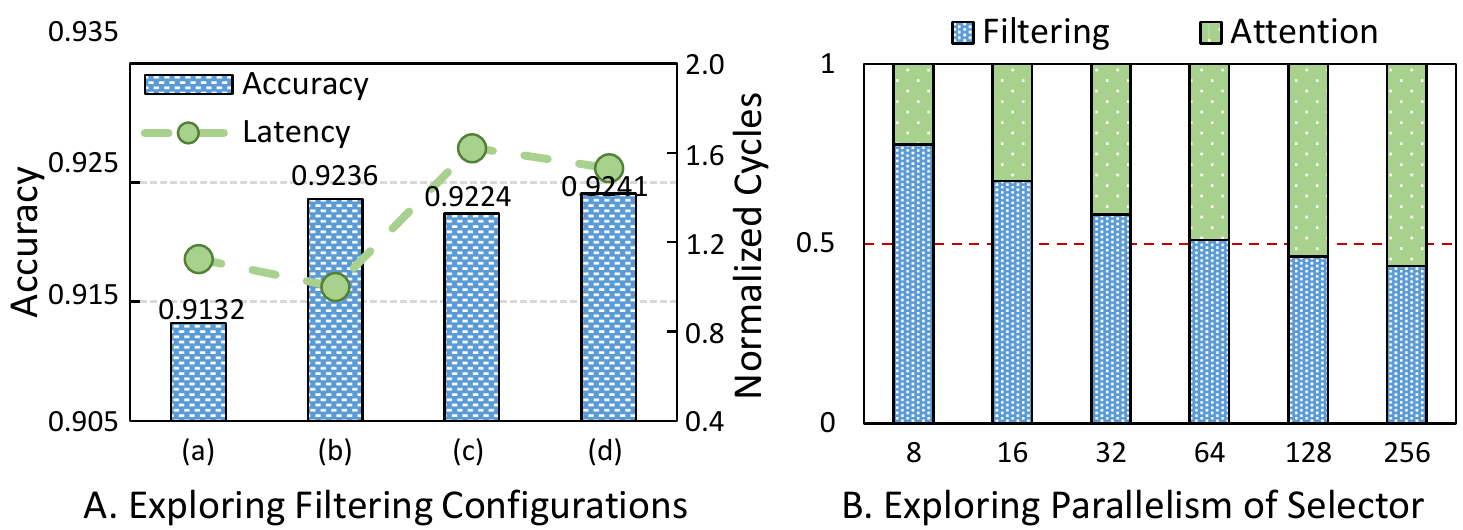} 
      \caption{Design Space Exploration. }
            \label{fig:explore DSE}
\end{figure}

\subsection{Design Space Exploration}
\textbf{Filtering Rounds:} 
\label{sec:round}
To investigate the most suitable filtering rounds, we choose Task-C as a representative task and test four configurations: (a) 1-2 (denotes using 1-bit tensors in the first round and 2-bit tensors in the second round), (b) 2-4, (c) 1-2-4 and (d) 2-4-8. We evaluate their accuracy on Task-C and estimate the average cycles spent by multi-round-filtering.  We adjust the pruning parameters of each configuration to make sure they have the same pruning ratio ($4 \times$) and the highest accuracy.  The comparison results are shown in Figure~\ref{fig:explore DSE}-A.  As we can see, among the four configurations, (b) 2-4 has the lowest execution cycles and high overall accuracy. Compared to (b) 2-4, (a) 1-2 has much lower accuracy, and since the 1-bit filtering is not accurate, it has a low pruning ratio in round-0, resulting in a longer total execution time than (b). The three-round filtering spends much more cycles than the 2-round filtering, and the (c) 1-2-4 has even lower accuracy because of using 1-bit filtering in round-0. The 2-4-8 filtering has slightly higher accuracy but demands 8-bit ALUs and takes more cycles.   Therefore, we argue that (b) 2-4 is an efficient choice for implementing multi-round-filtering. \rev{The top-k coverage analysis in Section \ref{sec:coverage} also proves that the 2-4 configuration achieves high top-k coverage on Task-A,C,D. Though it is not optimal for Task-B, for the current design, 2-4 configuration is the best selection for most of our tasks.}

\textbf{Parallelism of the Selector Module:}
Considering that the Selector module (see Section~\ref{sec:selector}) is in the critical path of FU, it will affect the overall throughput if its parallelism (the number of parallel comparators) is not sufficient. To determine a  suitable configuration, we explore a broad range of parallelism  (from 8 to 256) under the settings of Edge-server and test on Task-C. The ratios of total filtering cycles and attention cycles are shown in Figure~\ref{fig:explore DSE}-B. As we can see,  parallelism larger than 64 makes FU no longer the bottleneck, as FU and AU are pipelined during execution. Therefore, we adopt 64 comparators in a Selector to provide sufficient performance for multi-round filtering.


\section{\trev{Discussion}}

\label{sec:discussion}
\subsection{\trev{Full-Model Performance}} 

\trev{ The Energon co-processor is plugin-compatible with other NN accelerators, wherein only the attention operations are offloaded to Energon. To evaluate the full-model performance on such a heterogeneous system, we present a conceptional system and discuss its execution pipeline.}

\begin{figure} [t]
    \centering
    \includegraphics[width=1.0\linewidth]{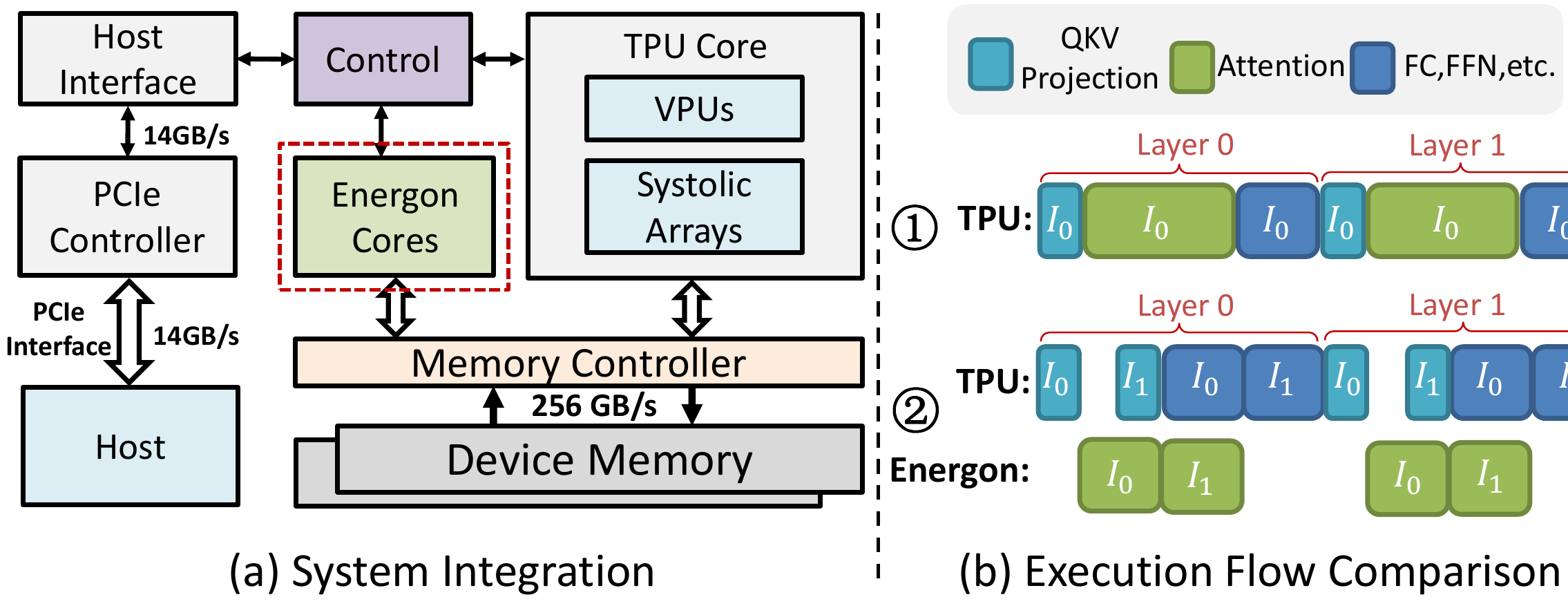} 
      \caption{\trev{(a) A conceptional integration of Energon with a TPU-like NN accelerator. (b) The execution flow comparison of \textcircled{{1}} TPU-only system and \textcircled{{2}} Energon-equipped system. }}
            \label{fig:full_system}
                  \vspace{-1em}
\end{figure}

\textbf{\trev{System Integration:}} 
\trev{As shown in  Figure~\ref{fig:full_system}-(a),
we demonstrate the integration of the Energon co-processor with a TPU-style \cite{tpu}  NN accelerator. The TPU core contains Vector-Processing-Units (VPUs) for vector operations and systolic arrays for matrix-matrix multiplication. Multiple Energon cores are placed beside the TPU core and share the same DRAM. With the shared device memory, Energon and TPU cores can access each other’s outputs without any extra data movement.  We list the system configurations in Table~\ref{tab:system_configurations}. We refer to PREMA~\cite{prema} for the setting of TPU parameters. We set eight Energon cores, which can process different queries in parallel. Both Energon and TPU cores execute operations according to the instructions sent from the host CPU~\cite{tpu}. } 

\trev{We compare the Energon-equipped system with the original TPU-only system. Figure~\ref{fig:full_system}-(b) shows their different execution flows. According to Figure~\ref{fig:transformer}, we divide the operations of transformer blocks into three sequentially-executed parts, namely the QKV projection operations, Attention operations and the remaining FC/FFN/Actication operations. In the Energon-equipped system, the Attention operations are executed on Energon, while the other two operations are executed by TPU. To improve the throughput, we enable pipelined execution in the Energon-equipped system. As Figure~\ref{fig:full_system}-(b) shows, the operations concerning two successive input sequences ($I_0$ and $I_1$) are interleaved on both TPU and Energon processors, greatly improving the system throughput. 
}

\begin{table}[t]
\centering
\caption{System Configurations}
\label{tab:system_configurations}
\renewcommand\arraystretch{1.1}
\setlength{\tabcolsep}{4mm}{
\resizebox{0.48\textwidth}{!}
{
\begin{tabular}{|cc|}
\hline
\multicolumn{2}{|c|}{\textbf{TPU Core Parameters}}                                \\ \hline
\multicolumn{1}{|c|}{Systolic-array dimension} & $128\times 128$                  \\ \hline
\multicolumn{1}{|c|}{VPU dimension}            & SIMD-16 $\times 32$              \\ \hline
\multicolumn{1}{|c|}{PE operating frequency}   & 700 MHz                          \\ \hline
\multicolumn{1}{|c|}{On-chip SRAM size}        & 8 (activations) + 4 (weights) MB \\ \hline
\multicolumn{2}{|c|}{\textbf{Energon Co-Processor Parameters:}}                   \\ \hline
\multicolumn{2}{|c|}{8 $\times$ Energon-server}                                    \\ \hline
\multicolumn{2}{|c|}{\textbf{Memory system}}                                      \\ \hline
\multicolumn{1}{|c|}{Memory bandwidth}         & 256GB/s                          \\ \hline
\multicolumn{1}{|c|}{Memory access latency}    & 100 cycles                       \\ \hline
\end{tabular}
}}
\end{table}

\begin{figure} [t]
    \centering
    \includegraphics[width=1.0\linewidth]{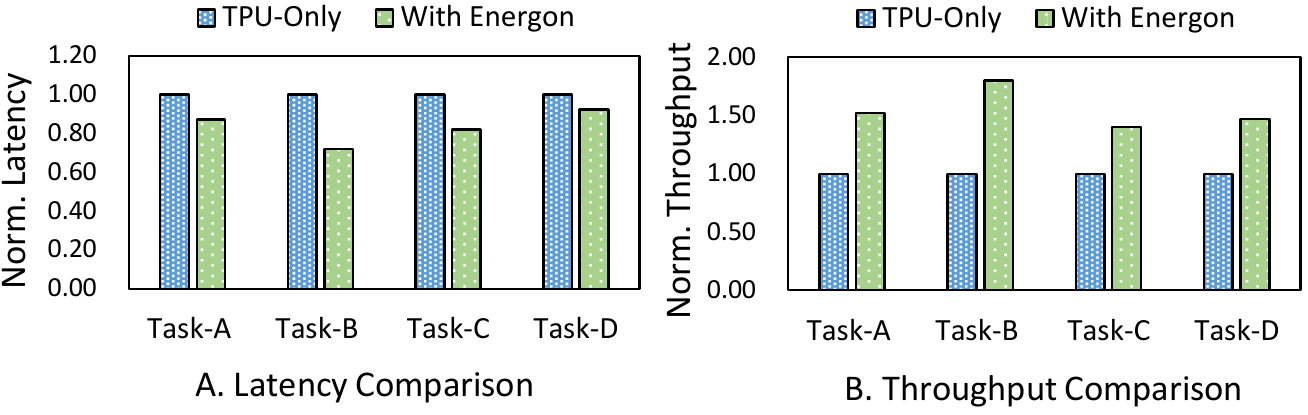} 
          \vspace{-1.5em}
      \caption{\trev{End-to-End performance comparison. } }
            \label{fig:e2e}
\end{figure}

\textbf{\trev{End-to-End Performance:}}
\trev{ We evaluate the full-network performance on both systems. To achieve this, we adopt  SCALE-Sim~\cite{scalesim} to simulate the behavior of the TPU core and integrate it with our Energon simulator. For fair comparisons, we scale up the frequency of the TPU-only system  and ensure that  the two systems have the same TOPS (Tera Operations Per Second).   Figure~\ref{fig:e2e} shows the end-to-end execution latency and throughput comparisons on the four tasks. With Energon, we can achieve $1.21\times$ lower end-to-end latency due to the saved attention cycles. Also, by offloading attention operations to Energon cores and enabling pipelined execution, Energon-equipped can achieve on-average $1.55\times$ higher throughput. In brief, Energon can obviously speed up the full-model network execution through dynamic sparse attention.}

\begin{figure} [t]
    \centering
    \includegraphics[width=0.96\linewidth]{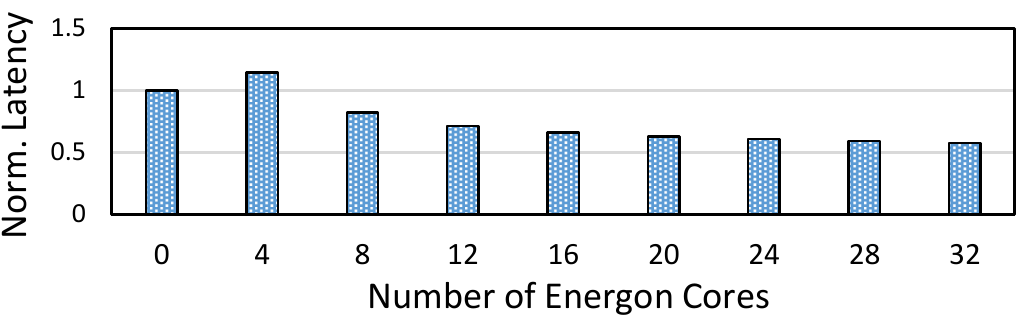} 
      \caption{\trev{Scalability exploration on Task-C. We use zero Energon cores to denote the TPU-only system. } }
            \label{fig:scalability}
\end{figure}

\textbf{\trev{Scalability Evaluation:}} \trev{By default, we equip eight Energon cores in the system. Equipping more cores may further reduce the attention cycles. Therefore,  we set different numbers of Energon cores and evaluate the scalability. As shown in Figure~\ref{fig:scalability}, four Energon cores cannot outperform the TPU-only system due to the insufficient computational capacity. Setting the core numbers from 4 to 16 reduces the end-to-end latency rapidly. However, keep adding the cores will not bring obvious performance gain. On the one hand, attention operations are no longer the bottleneck. On the other hand, all cores share the memory bandwidth. Too many cores may make the system memory-bounded, even with double buffers. }

\textbf{\trev{Comparison with SpAtten:}} \trev{SpAtten's cascaded token pruning algorithm can also prune fully-connected layers, which are unpruned in our settings.  However, as stated in Section~\ref{sec:comparison}, without model retraining, SpAtten's methods can only prune $2\times$ attention operations, while Energon can achieve $16\times,8\times,8\times$ pruning on Task-A,C,D. We  assume that SpAtten can also prune $2\times$ fully-connected layers and ignore  the huge accuracy loss (impossible though), then the non-attention operations of a SpAtten-equipped system should be roughly $2\times$ faster than our Energon-equipped system, if having the same GOPS. According to Section~\ref{sec:comparison},  Energon can provide  $1.7\times$ higher attention speed. Therefore, the two systems will achieve similar end-to-end performance in most cases where the cycles of attention and fully-connected layers are close. For attention-dominated cases, Energon will show more advantages.}

\section{Related work}

\label{sec:related}
\subsection{NN Accelerators}

There have been various accelerators designed for traditional DNNs. 
They accelerate convolution layers and fully-connected layers through 1-D inner-product units \cite{diannao,dadiannao,di_lstm_1} or 2-D PE arrays~\cite{tpu,shidiannao,eyeriss}.  These works leverage data-level parallelism to speed up the execution and also save energy through dataflow optimization. Some other works  \cite{eie,scnn,combricon-S,zhang2016cambricon,blockgnn,di_lstm_0}  exploit the sparsity of DNN models to reduce computation and eliminate memory footprint. Still, some NN accelerators leverage dynamic mix-precision to mitigate the computational complexity. Two recent works are DRQ~\cite{drq}, and  DUET \cite{DUET}. They adopt 4-bit tensors to compute the "unimportant" parts of the feature-maps and use 8-bit or 16-bit tensors to process the "important" parts, similar to our \trev{MP-}MRF mechanism. However, they are designed for convolutional layers and cannot directly support MP-MRF.

\subsection{Efficient Transformers}
Apart from the  sparse attention in Section~\ref{sec:sparse}, several recent works implement efficient transformers using  low-bit quantization, knowledge distillation, and even neural-architecture search (NAS), etc. To name a few, Q8Bert \cite{q8bert} adopts 8-bit quantization  to avoid costly floating-point operations.  GOBO~\cite{gobo} also proposes to compress the  majority of parameters to 3 bits while maintaining their accuracy.  QBert~\cite{qbert} and  TernaryBERT further quantize the model weights to  2 bits. HAT~\cite{HAT} leverages neural architecture search to design an efficient transformer model for edge devices. However, These low-bit transformers still need high-precision dense attention operations to maintain accuracy. Our proposed dynamic sparse attention mechanism can further reduce the computational overhead on top of these methods. 


\section{conclusion}
\label{sec:conclusion}
In this paper we propose Energon, a novel  algorithm-architecture co-design solution  to accelerate the emerging transformers with dynamic sparse attention. We present a \emph{Mix-Precision Multi-Round Filtering} (MP-MRF) algorithm to dynamically identify important query-key pairs at runtime using low-precision tensors and  only adopt high-precision tensors in the  attention stage to reduce the overall complexity. 
We also design an Energon co-processor architecture to accelerate such an algorithm. Elaborated pipelines and specialized  optimizations jointly boost the performance and reduce energy consumption.

\section*{Acknowledgment}
This work is supported by Key-Area Research and Development Program of Guangdong Province (2021B0101310002), NSF of China (61832020, 62032001, 92064006), Beijing Academy of Artificial Intelligence (BAAI), and 111 Project (B18001)


\ifCLASSOPTIONcaptionsoff
  \newpage
\fi

\bibliographystyle{IEEEtranS}
\bibliography{refs}

\begin{thebibliography}{10}
\providecommand{\url}[1]{#1}
\csname url@samestyle\endcsname
\providecommand{\newblock}{\relax}
\providecommand{\bibinfo}[2]{#2}
\providecommand{\BIBentrySTDinterwordspacing}{\spaceskip=0pt\relax}
\providecommand{\BIBentryALTinterwordstretchfactor}{4}
\providecommand{\BIBentryALTinterwordspacing}{\spaceskip=\fontdimen2\font plus
\BIBentryALTinterwordstretchfactor\fontdimen3\font minus
  \fontdimen4\font\relax}
\providecommand{\BIBforeignlanguage}[2]{{%
\expandafter\ifx\csname l@#1\endcsname\relax
\typeout{** WARNING: IEEEtranS.bst: No hyphenation pattern has been}%
\typeout{** loaded for the language `#1'. Using the pattern for}%
\typeout{** the default language instead.}%
\else
\language=\csname l@#1\endcsname
\fi
#2}}
\providecommand{\BIBdecl}{\relax}
\BIBdecl

\bibitem{hbminterface}
A.~Alzahmi, M.~Alswat, and C.-C. Lin, ``Energy efficient 5gb/s/pin baseband
  transceiver for 3d memory interface,'' in \emph{2020 10th Annual Computing
  and Communication Workshop and Conference (CCWC)}.\hskip 1em plus 0.5em minus
  0.4em\relax IEEE, 2020, pp. 0827--0829.

\bibitem{chisel}
J.~Bachrach, H.~Vo, B.~C. Richards, Y.~Lee, A.~Waterman, R.~Avizienis,
  J.~Wawrzynek, and K.~Asanovic, ``Chisel: constructing hardware in a scala
  embedded language,'' in \emph{The 49th Annual Design Automation Conference
  2012, {DAC} '12, San Francisco, CA, USA, June 3-7, 2012}, P.~Groeneveld,
  D.~Sciuto, and S.~Hassoun, Eds.\hskip 1em plus 0.5em minus 0.4em\relax {ACM},
  2012, pp. 1216--1225.

\bibitem{longformer}
I.~Beltagy, M.~E. Peters, and A.~Cohan, ``Longformer: The long-document
  transformer,'' \emph{CoRR}, vol. abs/2004.05150, 2020.

\bibitem{detr}
N.~Carion, F.~Massa, G.~Synnaeve, N.~Usunier, A.~Kirillov, and S.~Zagoruyko,
  ``End-to-end object detection with transformers,'' \emph{arXiv preprint
  arXiv:2005.12872}, 2020.

\bibitem{chandrasekar2012drampower}
K.~Chandrasekar, C.~Weis, Y.~Li, B.~Akesson, N.~Wehn, and K.~Goossens,
  ``Drampower: Open-source dram power \& energy estimation tool,'' \emph{URL:
  http://www. drampower. info}, vol.~22, 2012.

\bibitem{diannao}
T.~Chen, Z.~Du, N.~Sun, J.~Wang, C.~Wu, Y.~Chen, and O.~Temam, ``Diannao: a
  small-footprint high-throughput accelerator for ubiquitous
  machine-learning,'' in \emph{Architectural Support for Programming Languages
  and Operating Systems, {ASPLOS} '14, Salt Lake City, UT, USA, March 1-5,
  2014}, R.~Balasubramonian, A.~Davis, and S.~V. Adve, Eds.\hskip 1em plus
  0.5em minus 0.4em\relax {ACM}, 2014, pp. 269--284.

\bibitem{eyeriss}
Y.~Chen, J.~S. Emer, and V.~Sze, ``Eyeriss: {A} spatial architecture for
  energy-efficient dataflow for convolutional neural networks,'' in \emph{43rd
  {ACM/IEEE} Annual International Symposium on Computer Architecture, {ISCA}
  2016, Seoul, South Korea, June 18-22, 2016}.\hskip 1em plus 0.5em minus
  0.4em\relax {IEEE} Computer Society, 2016, pp. 367--379.

\bibitem{dadiannao}
Y.~Chen, T.~Luo, S.~Liu, S.~Zhang, L.~He, J.~Wang, L.~Li, T.~Chen, Z.~Xu,
  N.~Sun, and O.~Temam, ``Dadiannao: {A} machine-learning supercomputer,'' in
  \emph{47th Annual {IEEE/ACM} International Symposium on Microarchitecture,
  {MICRO}}.\hskip 1em plus 0.5em minus 0.4em\relax {IEEE} Computer Society,
  2014, pp. 609--622.

\bibitem{sparse}
R.~Child, S.~Gray, A.~Radford, and I.~Sutskever, ``Generating long sequences
  with sparse transformers,'' \emph{arXiv preprint arXiv:1904.10509}, 2019.

\bibitem{prema}
Y.~Choi and M.~Rhu, ``Prema: A predictive multi-task scheduling algorithm for
  preemptible neural processing units,'' in \emph{2020 IEEE International
  Symposium on High Performance Computer Architecture (HPCA)}.\hskip 1em plus
  0.5em minus 0.4em\relax IEEE, 2020, pp. 220--233.

\bibitem{imagenet}
J.~{Deng}, W.~{Dong}, R.~{Socher}, L.~{Li}, {Kai Li}, and {Li Fei-Fei},
  ``Imagenet: A large-scale hierarchical image database,'' in \emph{2009 IEEE
  Conference on Computer Vision and Pattern Recognition}, 2009, pp. 248--255.

\bibitem{bert}
J.~Devlin, M.-W. Chang, K.~Lee, and K.~Toutanova, ``Bert: Pre-training of deep
  bidirectional transformers for language understanding,'' \emph{arXiv preprint
  arXiv:1810.04805}, 2018.

\bibitem{vit}
A.~Dosovitskiy, L.~Beyer, A.~Kolesnikov, D.~Weissenborn, X.~Zhai,
  T.~Unterthiner, M.~Dehghani, M.~Minderer, G.~Heigold, S.~Gelly \emph{et~al.},
  ``An image is worth 16x16 words: Transformers for image recognition at
  scale,'' \emph{arXiv preprint arXiv:2010.11929}, 2020.

\bibitem{shidiannao}
Z.~Du, R.~Fasthuber, T.~Chen, P.~Ienne, L.~Li, T.~Luo, X.~Feng, Y.~Chen, and
  O.~Temam, ``Shidiannao: Shifting vision processing closer to the sensor,''
  \emph{SIGARCH Comput. Archit. News}, vol.~43, no.~3S, p. 92–104, Jun. 2015.

\bibitem{a3}
T.~J. Ham, S.~J. Jung, S.~Kim, Y.~H. Oh, Y.~Park, Y.~Song, J.-H. Park, S.~Lee,
  K.~Park, J.~W. Lee \emph{et~al.}, ``A\^{} 3: Accelerating attention
  mechanisms in neural networks with approximation,'' in \emph{2020 IEEE
  International Symposium on High Performance Computer Architecture
  (HPCA)}.\hskip 1em plus 0.5em minus 0.4em\relax IEEE, 2020, pp. 328--341.

\bibitem{eie}
S.~Han, X.~Liu, H.~Mao, J.~Pu, A.~Pedram, M.~A. Horowitz, and W.~J. Dally,
  ``{EIE:} efficient inference engine on compressed deep neural network,'' in
  \emph{43rd {ACM/IEEE} Annual International Symposium on Computer
  Architecture, {ISCA} 2016, Seoul, South Korea, June 18-22, 2016}.\hskip 1em
  plus 0.5em minus 0.4em\relax {IEEE} Computer Society, 2016, pp. 243--254.

\bibitem{deepcompression}
S.~Han, H.~Mao, and W.~J. Dally, ``Deep compression: Compressing deep neural
  networks with pruning, trained quantization and huffman coding,'' \emph{arXiv
  preprint arXiv:1510.00149}, 2015.

\bibitem{gelu}
D.~Hendrycks and K.~Gimpel, ``Gaussian error linear units (gelus),''
  \emph{arXiv preprint arXiv:1606.08415}, 2016.

\bibitem{ho2019axial}
J.~Ho, N.~Kalchbrenner, D.~Weissenborn, and T.~Salimans, ``Axial attention in
  multidimensional transformers,'' \emph{arXiv preprint arXiv:1912.12180},
  2019.

\bibitem{hugging_face}
huggingface, ``Huggingface transformers,''
  \url{https://github.com/huggingface/transformers}.

\bibitem{XEON}
Intel, ``Intel xeon gold 5220 processor,'' \url{
  https://ark.intel.com/content/www/us/en/ark/products/193388/intel-xeon-gold-5220-processor-24-75m-cache-2-20-ghz.html}.

\bibitem{pytorch_vit}
jeonsworld, ``Vit-pytorch,'' \url{https://github.com/jeonsworld/ViT-pytorch }.

\bibitem{tpu}
N.~P. Jouppi, C.~Young, N.~Patil, D.~Patterson, G.~Agrawal, R.~Bajwa, S.~Bates,
  S.~Bhatia, N.~Boden, A.~Borchers \emph{et~al.}, ``In-datacenter performance
  analysis of a tensor processing unit,'' in \emph{Proceedings of the 44th
  annual international symposium on computer architecture}, 2017, pp. 1--12.

\bibitem{ramulator}
\BIBentryALTinterwordspacing
Y.~Kim, W.~Yang, and O.~Mutlu, ``Ramulator: {A} fast and extensible {DRAM}
  simulator,'' \emph{{IEEE} Comput. Archit. Lett.}, vol.~15, no.~1, pp. 45--49,
  2016. [Online]. Available: \url{https://doi.org/10.1109/LCA.2015.2414456}
\BIBentrySTDinterwordspacing

\bibitem{reformer}
N.~Kitaev, L.~Kaiser, and A.~Levskaya, ``Reformer: The efficient transformer,''
  in \emph{8th International Conference on Learning Representations, {ICLR}
  2020, Addis Ababa, Ethiopia, April 26-30, 2020}.\hskip 1em plus 0.5em minus
  0.4em\relax OpenReview.net, 2020.

\bibitem{cifar}
A.~Krizhevsky, G.~Hinton \emph{et~al.}, ``Learning multiple layers of features
  from tiny images,'' 2009.

\bibitem{alexnet}
A.~Krizhevsky, I.~Sutskever, and G.~E. Hinton, ``Imagenet classification with
  deep convolutional neural networks,'' in \emph{Advances in Neural Information
  Processing Systems 25: 26th Annual Conference on Neural Information
  Processing Systems,{NIPS}}, P.~L. Bartlett, F.~C.~N. Pereira, C.~J.~C.
  Burges, L.~Bottou, and K.~Q. Weinberger, Eds., 2012, pp. 1106--1114.

\bibitem{albert}
Z.~Lan, M.~Chen, S.~Goodman, K.~Gimpel, P.~Sharma, and R.~Soricut, ``Albert: A
  lite bert for self-supervised learning of language representations,''
  \emph{arXiv preprint arXiv:1909.11942}, 2019.

\bibitem{bart}
M.~Lewis, Y.~Liu, N.~Goyal, M.~Ghazvininejad, A.~Mohamed, O.~Levy, V.~Stoyanov,
  and L.~Zettlemoyer, ``{BART:} denoising sequence-to-sequence pre-training for
  natural language generation, translation, and comprehension,'' in
  \emph{Proceedings of the 58th Annual Meeting of the Association for
  Computational Linguistics, {ACL} 2020, Online, July 5-10, 2020}, D.~Jurafsky,
  J.~Chai, N.~Schluter, and J.~R. Tetreault, Eds.\hskip 1em plus 0.5em minus
  0.4em\relax Association for Computational Linguistics, 2020, pp. 7871--7880.

\bibitem{DUET}
L.~Liu, Z.~Qu, L.~Deng, F.~Tu, S.~Li, X.~Hu, Z.~Gu, Y.~Ding, and Y.~Xie,
  ``Duet: Boosting deep neural network efficiency on dual-module
  architecture,'' in \emph{2020 53rd Annual IEEE/ACM International Symposium on
  Microarchitecture (MICRO)}.\hskip 1em plus 0.5em minus 0.4em\relax IEEE,
  2020, pp. 738--750.

\bibitem{roberta}
Y.~Liu, M.~Ott, N.~Goyal, J.~Du, M.~Joshi, D.~Chen, O.~Levy, M.~Lewis,
  L.~Zettlemoyer, and V.~Stoyanov, ``Roberta: A robustly optimized bert
  pretraining approach,'' \emph{arXiv preprint arXiv:1907.11692}, 2019.

\bibitem{wikitext}
S.~Merity, C.~Xiong, J.~Bradbury, and R.~Socher, ``Pointer sentinel mixture
  models,'' in \emph{5th International Conference on Learning Representations,
  {ICLR} 2017, Toulon, France, April 24-26, 2017, Conference Track
  Proceedings}.\hskip 1em plus 0.5em minus 0.4em\relax OpenReview.net, 2017.

\bibitem{LPDDR3}
micron, ``8gb, 16gb: 253-ball, dual-channel mobile lpddr3 sdram,'' \url{
  https://www.micron.com/-/media/client/global/documents/products/data-sheet/dram/mobile-dram/low-power-dram/lpddr3/253b_12-5x12-5_2ch_8-16gb_2c0f_mobile_lpddr3.pdf?rev=1b66d5710434460eb13dc3be8faa6d77}.

\bibitem{rnn}
T.~Mikolov, M.~Karafi{\'a}t, L.~Burget, J.~{\v{C}}ernock{\`y}, and
  S.~Khudanpur, ``Recurrent neural network based language model,'' in
  \emph{Eleventh annual conference of the international speech communication
  association}, 2010.

\bibitem{freePDK}
NCSU, ``Freepdk45,'' \url{https://www.eda.ncsu.edu/wiki/FreePDK45:Contents}.

\bibitem{di_lstm_0}
S.~Nelson, W.~Khalil, S.~Kim, J.~Di, Z.~Zhou, Z.~Yuan, and G.~Sun, ``Rapid
  configuration of asynchronous recurrent neural networks for asic
  implementations,'' in \emph{2021 IEEE High Performance Extreme Computing
  Conference (HPEC)}, 2021, pp. 1--6.

\bibitem{di_lstm_1}
S.~Nelson, S.~Y. Kim, J.~Di, Z.~Zhou, Z.~Yuan, and G.~Sun, ``Reconfigurable
  asic implementation of asynchronous recurrent neural networks,'' in
  \emph{2021 27th IEEE International Symposium on Asynchronous Circuits and
  Systems (ASYNC)}, 2021, pp. 48--54.

\bibitem{softmax}
P.~Nilsson, A.~U.~R. Shaik, R.~Gangarajaiah, and E.~Hertz, ``Hardware
  implementation of the exponential function using taylor series,'' in
  \emph{2014 NORCHIP, Tampere, Finland, October 27-28, 2014}.\hskip 1em plus
  0.5em minus 0.4em\relax {IEEE}, 2014, pp. 1--4.

\bibitem{TX2}
NVIDIA, ``Nvidia jetson tx2,'' \url{
  https://www.nvidia.com/en-us/autonomous-machines/embedded-systems/jetson-tx2/}.

\bibitem{V100}
------, ``Nvidia v100 tensor core gpu,'' \url{
  https://www.nvidia.com/en-us/data-center/v100}.

\bibitem{cacti}
H.~Packard, ``Cacti,'' \url{ https://github.com/HewlettPackard/cacti.git}.

\bibitem{lpddr3Controller}
R.~Palmer, J.~Poulton, B.~Leibowitz, Y.~Frans, and N.~Nguyen, ``A 4.3gb/s
  mobile memory interface with power-efficient bandwidth scaling,'' in
  \emph{Symposium on Vlsi Circuits}, 2009.

\bibitem{scnn}
A.~Parashar, M.~Rhu, A.~Mukkara, A.~Puglielli, R.~Venkatesan, B.~Khailany,
  J.~S. Emer, S.~W. Keckler, and W.~J. Dally, ``{SCNN:} an accelerator for
  compressed-sparse convolutional neural networks,'' in \emph{Proceedings of
  the 44th Annual International Symposium on Computer Architecture,
  {ISCA}}.\hskip 1em plus 0.5em minus 0.4em\relax {ACM}, 2017, pp. 27--40.

\bibitem{pytorch}
A.~Paszke, S.~Gross, F.~Massa, A.~Lerer, J.~Bradbury, G.~Chanan, T.~Killeen,
  Z.~Lin, N.~Gimelshein, L.~Antiga \emph{et~al.}, ``Pytorch: An imperative
  style, high-performance deep learning library,'' \emph{arXiv preprint
  arXiv:1912.01703}, 2019.

\bibitem{PyRAPL}
powerapi ng, ``pyrapl,'' \url{ https://github.com/powerapi-ng/pyRAPL}.

\bibitem{blockbert}
J.~Qiu, H.~Ma, O.~Levy, W.~Yih, S.~Wang, and J.~Tang, ``Blockwise
  self-attention for long document understanding,'' in \emph{Proceedings of the
  2020 Conference on Empirical Methods in Natural Language Processing:
  Findings, {EMNLP} 2020, Online Event, 16-20 November 2020}, T.~Cohn, Y.~He,
  and Y.~Liu, Eds.\hskip 1em plus 0.5em minus 0.4em\relax Association for
  Computational Linguistics, 2020, pp. 2555--2565.

\bibitem{gpt2}
A.~Radford, J.~Wu, R.~Child, D.~Luan, D.~Amodei, and I.~Sutskever, ``Language
  models are unsupervised multitask learners,'' \emph{OpenAI blog}, vol.~1,
  no.~8, p.~9, 2019.

\bibitem{t5}
C.~Raffel, N.~Shazeer, A.~Roberts, K.~Lee, S.~Narang, M.~Matena, Y.~Zhou,
  W.~Li, and P.~J. Liu, ``Exploring the limits of transfer learning with a
  unified text-to-text transformer,'' \emph{J. Mach. Learn. Res.}

\bibitem{squad}
P.~Rajpurkar, J.~Zhang, K.~Lopyrev, and P.~Liang, ``Squad: 100, 000+ questions
  for machine comprehension of text,'' in \emph{Proceedings of the 2016
  Conference on Empirical Methods in Natural Language Processing, {EMNLP} 2016,
  Austin, Texas, USA, November 1-4, 2016}, J.~Su, X.~Carreras, and K.~Duh,
  Eds.\hskip 1em plus 0.5em minus 0.4em\relax The Association for Computational
  Linguistics, 2016, pp. 2383--2392.

\bibitem{raspberry}
Raspberry, ``Raspberry pi 4 computer model b,'' \url{
  https://static.raspberrypi.org/files/product-briefs/Raspberry-Pi-4-Product-Brief.pdf}.

\bibitem{PyNVML}
Rjzamora, ``pyrapl,'' \url{ https://pypi.org/project/pynvml}.

\bibitem{routing_transformer}
\BIBentryALTinterwordspacing
A.~Roy, M.~Saffar, A.~Vaswani, and D.~Grangier, ``Efficient content-based
  sparse attention with routing transformers,'' \emph{CoRR}, vol.
  abs/2003.05997, 2020. [Online]. Available:
  \url{https://arxiv.org/abs/2003.05997}
\BIBentrySTDinterwordspacing

\bibitem{scalesim}
A.~Samajdar, Y.~Zhu, P.~Whatmough, M.~Mattina, and T.~Krishna, ``Scale-sim:
  Systolic cnn accelerator simulator,'' \emph{arXiv preprint arXiv:1811.02883},
  2018.

\bibitem{video_transformer}
H.~Seong, J.~Hyun, and E.~Kim, ``Video multitask transformer network,'' in
  \emph{Proceedings of the IEEE/CVF International Conference on Computer Vision
  Workshops}, 2019, pp. 0--0.

\bibitem{qbert}
S.~Shen, Z.~Dong, J.~Ye, L.~Ma, Z.~Yao, A.~Gholami, M.~W. Mahoney, and
  K.~Keutzer, ``{Q-BERT:} hessian based ultra low precision quantization of
  {BERT},'' in \emph{The Thirty-Fourth {AAAI} Conference on Artificial
  Intelligence, {AAAI} 2020, The Thirty-Second Innovative Applications of
  Artificial Intelligence Conference, {IAAI} 2020, The Tenth {AAAI} Symposium
  on Educational Advances in Artificial Intelligence, {EAAI} 2020, New York,
  NY, USA, February 7-12, 2020}.\hskip 1em plus 0.5em minus 0.4em\relax {AAAI}
  Press, 2020, pp. 8815--8821.

\bibitem{vgg}
K.~Simonyan and A.~Zisserman, ``Very deep convolutional networks for
  large-scale image recognition,'' in \emph{3rd International Conference on
  Learning Representations, {ICLR}}, Y.~Bengio and Y.~LeCun, Eds., 2015.

\bibitem{snyder2004verilator}
W.~Snyder, ``Verilator and {SystemPerl},'' in \emph{North American SystemC
  Users' Group (NASCUG) Meeting at Design Automation Conference}, 2004.

\bibitem{sst2}
R.~Socher, A.~Perelygin, J.~Wu, J.~Chuang, C.~Manning, A.~Ng, and C.~Potts,
  ``{Parsing With Compositional Vector Grammars},'' in \emph{{EMNLP}}, 2013.

\bibitem{drq}
Z.~Song, B.~Fu, F.~Wu, Z.~Jiang, L.~Jiang, N.~Jing, and X.~Liang, ``Drq:
  dynamic region-based quantization for deep neural network acceleration,'' in
  \emph{2020 ACM/IEEE 47th Annual International Symposium on Computer
  Architecture (ISCA)}.\hskip 1em plus 0.5em minus 0.4em\relax IEEE, 2020, pp.
  1010--1021.

\bibitem{botnet}
A.~Srinivas, T.-Y. Lin, N.~Parmar, J.~Shlens, P.~Abbeel, and A.~Vaswani,
  ``Bottleneck transformers for visual recognition,'' \emph{arXiv preprint
  arXiv:2101.11605}, 2021.

\bibitem{videobert}
C.~Sun, A.~Myers, C.~Vondrick, K.~Murphy, and C.~Schmid, ``Videobert: A joint
  model for video and language representation learning,'' in \emph{Proceedings
  of the IEEE/CVF International Conference on Computer Vision (ICCV)}, October
  2019.

\bibitem{Sinkhorn}
Y.~Tay, D.~Bahri, L.~Yang, D.~Metzler, and D.~Juan, ``Sparse sinkhorn
  attention,'' in \emph{Proceedings of the 37th International Conference on
  Machine Learning, {ICML} 2020, 13-18 July 2020, Virtual Event}, ser.
  Proceedings of Machine Learning Research, vol. 119.\hskip 1em plus 0.5em
  minus 0.4em\relax {PMLR}, 2020, pp. 9438--9447.

\bibitem{efficient}
Y.~Tay, M.~Dehghani, D.~Bahri, and D.~Metzler, ``Efficient transformers: A
  survey,'' \emph{arXiv preprint arXiv:2009.06732}, 2020.

\bibitem{deit}
H.~Touvron, M.~Cord, M.~Douze, F.~Massa, A.~Sablayrolles, and H.~J{\'e}gou,
  ``Training data-efficient image transformers \& distillation through
  attention,'' \emph{arXiv preprint arXiv:2012.12877}, 2020.

\bibitem{transformer}
A.~Vaswani, N.~Shazeer, N.~Parmar, J.~Uszkoreit, L.~Jones, A.~N. Gomez,
  {\L}.~Kaiser, and I.~Polosukhin, ``Attention is all you need,'' in
  \emph{Advances in neural information processing systems}, 2017, pp.
  5998--6008.

\bibitem{GLUE}
A.~Wang, A.~Singh, J.~Michael, F.~Hill, O.~Levy, and S.~R. Bowman, ``{GLUE:}
  {A} multi-task benchmark and analysis platform for natural language
  understanding,'' in \emph{7th International Conference on Learning
  Representations, {ICLR} 2019, New Orleans, LA, USA, May 6-9, 2019}.\hskip 1em
  plus 0.5em minus 0.4em\relax OpenReview.net, 2019.

\bibitem{HAT}
H.~Wang, Z.~Wu, Z.~Liu, H.~Cai, L.~Zhu, C.~Gan, and S.~Han, ``{HAT:}
  hardware-aware transformers for efficient natural language processing,'' in
  \emph{Proceedings of the 58th Annual Meeting of the Association for
  Computational Linguistics, {ACL} 2020, Online, July 5-10, 2020}, D.~Jurafsky,
  J.~Chai, N.~Schluter, and J.~R. Tetreault, Eds.\hskip 1em plus 0.5em minus
  0.4em\relax Association for Computational Linguistics, 2020, pp. 7675--7688.

\bibitem{spatten}
H.~Wang, Z.~Zhang, and S.~Han, ``Spatten: Efficient sparse attention
  architecture with cascade token and head pruning,'' \emph{arXiv preprint
  arXiv:2012.09852}, 2020.

\bibitem{hbm}
Xilinx, ``Virtex ultrascale+ hbm fpga: a revolutionary increase in memory
  performance,'' \url{
  https://www.xilinx.com/support/documentation/white_papers/wp485-hbm.pdf}.

\bibitem{xlnet}
Z.~Yang, Z.~Dai, Y.~Yang, J.~Carbonell, R.~R. Salakhutdinov, and Q.~V. Le,
  ``Xlnet: Generalized autoregressive pretraining for language understanding,''
  in \emph{Advances in neural information processing systems}, 2019, pp.
  5753--5763.

\bibitem{gobo}
A.~H. Zadeh, I.~Edo, O.~M. Awad, and A.~Moshovos, ``Gobo: Quantizing
  attention-based nlp models for low latency and energy efficient inference,''
  in \emph{2020 53rd Annual IEEE/ACM International Symposium on
  Microarchitecture (MICRO)}.\hskip 1em plus 0.5em minus 0.4em\relax IEEE,
  2020, pp. 811--824.

\bibitem{q8bert}
O.~Zafrir, G.~Boudoukh, P.~Izsak, and M.~Wasserblat, ``Q8bert: Quantized 8bit
  bert,'' \emph{arXiv preprint arXiv:1910.06188}, 2019.

\bibitem{zhang2016cambricon}
S.~Zhang, Z.~Du, L.~Zhang, H.~Lan, S.~Liu, L.~Li, Q.~Guo, T.~Chen, and Y.~Chen,
  ``Cambricon-x: An accelerator for sparse neural networks,'' in \emph{2016
  49th Annual IEEE/ACM International Symposium on Microarchitecture
  (MICRO)}.\hskip 1em plus 0.5em minus 0.4em\relax IEEE, 2016, pp. 1--12.

\bibitem{ternaryBert}
W.~Zhang, L.~Hou, Y.~Yin, L.~Shang, X.~Chen, X.~Jiang, and Q.~Liu,
  ``Ternarybert: Distillation-aware ultra-low bit {BERT},'' in
  \emph{Proceedings of the 2020 Conference on Empirical Methods in Natural
  Language Processing, {EMNLP} 2020, Online, November 16-20, 2020}, B.~Webber,
  T.~Cohn, Y.~He, and Y.~Liu, Eds.\hskip 1em plus 0.5em minus 0.4em\relax
  Association for Computational Linguistics, 2020, pp. 509--521.

\bibitem{informer}
H.~Zhou, S.~Zhang, J.~Peng, S.~Zhang, J.~Li, H.~Xiong, and W.~Zhang,
  ``Informer: Beyond efficient transformer for long sequence time-series
  forecasting,'' \emph{CoRR}, vol. abs/2012.07436, 2020.

\bibitem{combricon-S}
X.~Zhou, Z.~Du, Q.~Guo, S.~Liu, C.~Liu, C.~Wang, X.~Zhou, L.~Li, T.~Chen, and
  Y.~Chen, ``Cambricon-s: Addressing irregularity in sparse neural networks
  through {A} cooperative software/hardware approach,'' in \emph{51st Annual
  {IEEE/ACM} International Symposium on Microarchitecture, {MICRO}}.\hskip 1em
  plus 0.5em minus 0.4em\relax {IEEE} Computer Society, 2018, pp. 15--28.

\bibitem{blockgnn}
Z.~Zhou, B.~Shi, Z.~Zhang, Y.~Guan, G.~Sun, and G.~Luo, ``Blockgnn: Towards
  efficient gnn acceleration using block-circulant weight matrices,'' in
  \emph{2021 58th ACM/IEEE Design Automation Conference (DAC)}.\hskip 1em plus
  0.5em minus 0.4em\relax IEEE, 2021, pp. 1009--1014.

\bibitem{deformable}
X.~Zhu, W.~Su, L.~Lu, B.~Li, X.~Wang, and J.~Dai, ``Deformable detr: Deformable
  transformers for end-to-end object detection,'' \emph{arXiv preprint
  arXiv:2010.04159}, 2020.

\end{thebibliography}




\vspace{-2em}

\begin{IEEEbiography}[{\includegraphics[width=1in,height=1.25in,clip,keepaspectratio]{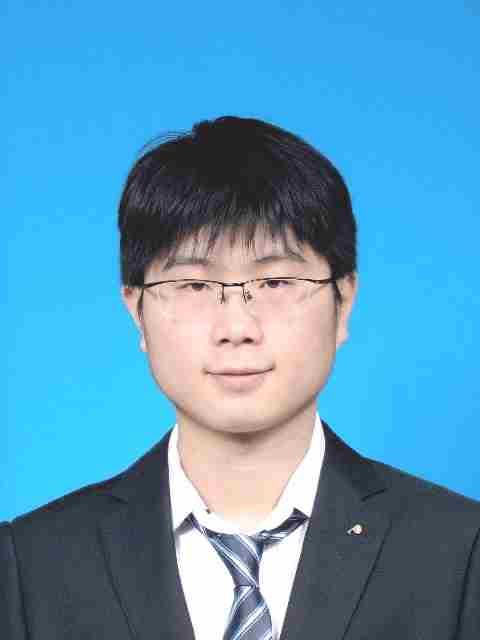}}]{Zhe Zhou}
received the BS degree from Peking University, Beijing, China, in 2019. He is currently working toward the PhD degree in computer science at Peking University, China. His current research interests include computer architecture,  domain-specific accelerator, and near-memory processing technology.\end{IEEEbiography}
\vspace{-2em}

\begin{IEEEbiography}[{\includegraphics[width=1in,height=1.25in,clip,keepaspectratio]{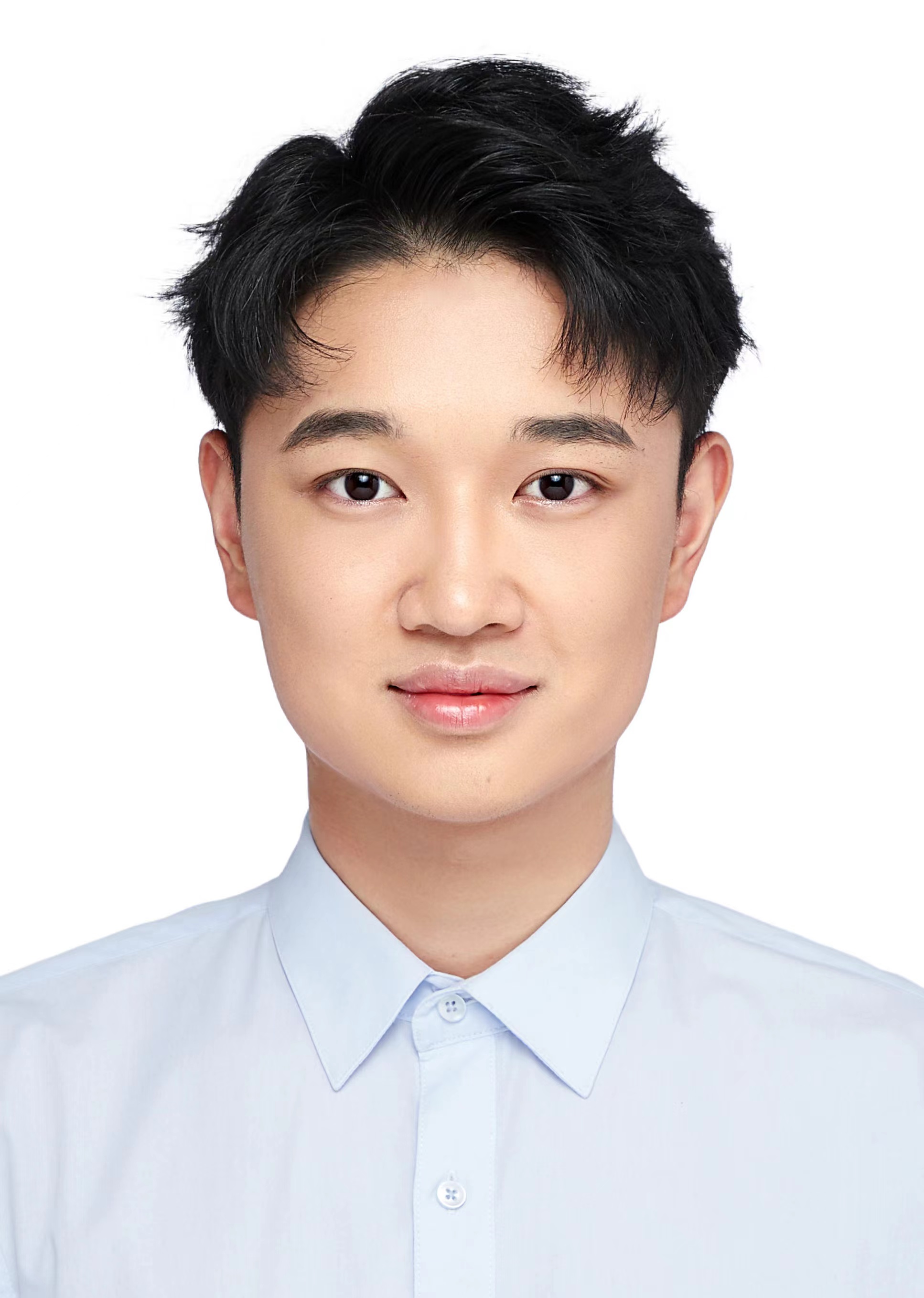}}]{Junlin Liu}
 is a senior bachelor student of EECS at Peking University, major in computer science. His current research interests include machine learning systems and distributed computing.
\end{IEEEbiography}
\vspace{-2em}

\begin{IEEEbiography}[{\includegraphics[width=1in,height=1.25in,clip,keepaspectratio]{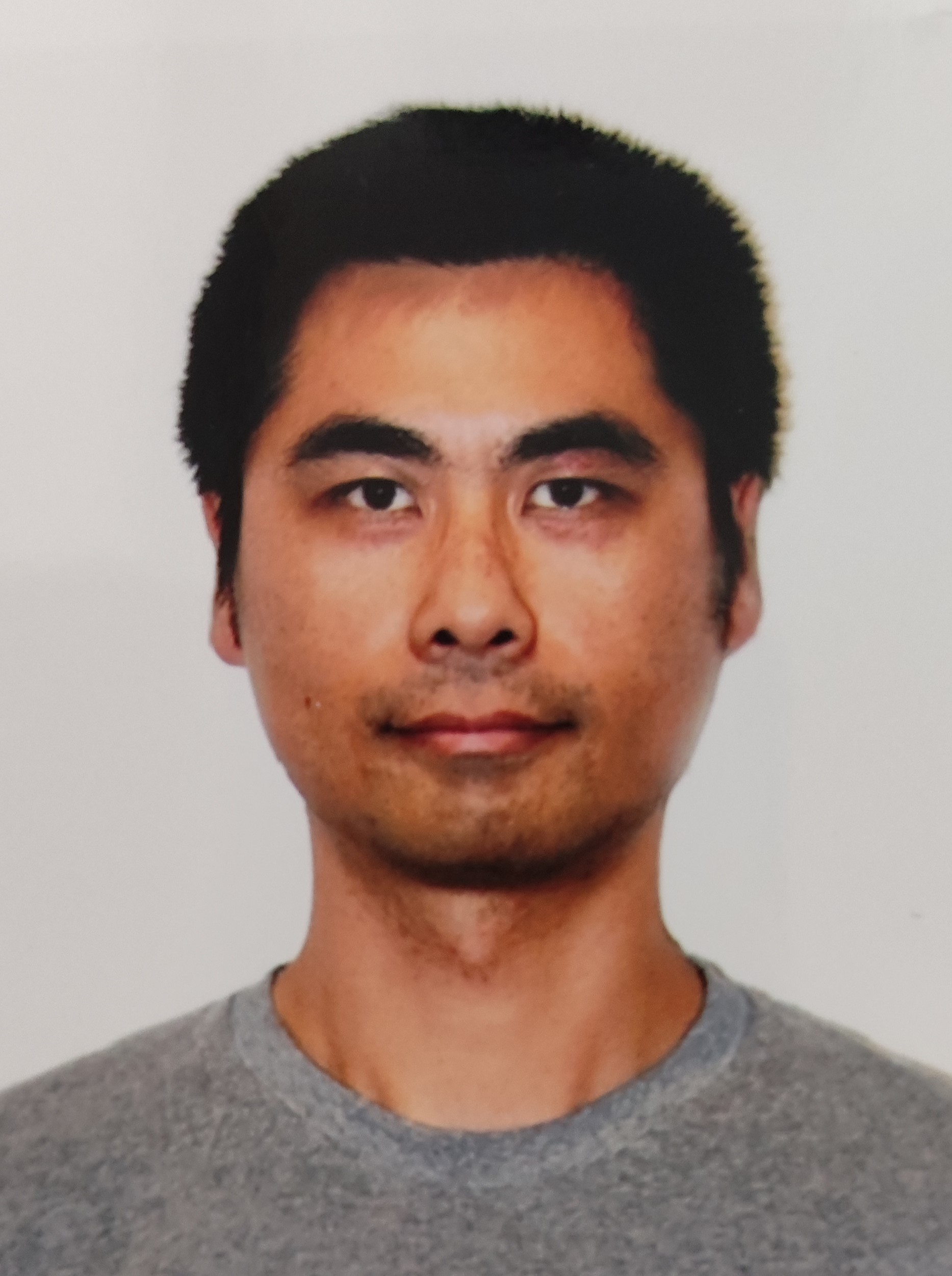}}]{Zhenyu Gu}
Zhenyu GU received his BS and MS from Fudan university, shanghai, China in 2000 and 2003, respectively. He got his PhD degree from EECS department of  Northwestern University in 2007. His research interests include computer architecture, hardware software co-design and electronic design automation.
\end{IEEEbiography}
\vspace{-2em}

\begin{IEEEbiography}[{\includegraphics[width=1in,height=1.25in,clip,keepaspectratio]{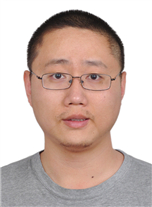}}]{Guangyu Sun}
is an associate professor of Center for Energy-efficient Computing and Applications (CECA) at Peking University. He received his B.S. and M.S degrees from Tsinghua University, Beijing, in 2003 and 2006, respectively. He received his Ph.D. degree in Computer Science from the Pennsylvania State University in 2011. His research interests include computer architecture, acceleration system, and electronic design automation for modern applications. Dr. Sun is now serving as an AE of ACM JETC. He is a member of IEEE, ACM, and CCF.
\end{IEEEbiography}


\end{document}